\documentclass[twocolumn]{aastex63}
\usepackage{graphicx}
\usepackage{color}
\usepackage{amsmath,amssymb}
\usepackage{gensymb}
\usepackage{times}
\usepackage{nicefrac}
\newcommand{\kms}{km s$^{-1}$}

\newcommand{\mstar}{$M_{\star}$}
\newcommand{\msun}{$M_{\odot}$}
\newcommand{\feh}{$[\rm{Fe}/\rm{H}]$}
\newcommand{\afe}{$[\alpha /\rm{Fe}]~$}

\usepackage[utf8]{inputenc}
\begin{document}
\title{Reading the CARDs: the Imprint of Accretion History in the Chemical Abundances of the Milky Way's Stellar Halo}
\author[0000-0002-6993-0826]{Emily C. Cunningham}
\affiliation{Center for Computational Astrophysics, Flatiron Institute, 162 5th Ave., New York, NY 10010, USA}
\author[0000-0003-3939-3297]{Robyn E. Sanderson}
\affiliation{Department of Physics and Astronomy, University of Pennsylvania, 209 S 33rd St., Philadelphia, PA 19104, USA}
\affiliation{Center for Computational Astrophysics, Flatiron Institute, 162 5th Ave., New York, NY 10010, USA}
\author[0000-0001-6244-6727]{Kathryn V. Johnston}
\affiliation{Department of Astronomy, Columbia University, 550 West 120th Street, New York, NY, 10027, USA}
\affiliation{Center for Computational Astrophysics, Flatiron Institute, 162 5th Ave., New York, NY 10010, USA}
\author[0000-0001-5214-8822]{Nondh Panithanpaisal}
\affil{Department of Physics and Astronomy, University of Pennsylvania, 209 S 33rd St., Philadelphia, PA 19104, USA}
\author[0000-0001-5082-6693]{Melissa K. Ness}
\affiliation{Department of Astronomy, Columbia University, 550 West 120th Street, New York, NY, 10027, USA}
\affiliation{Center for Computational Astrophysics, Flatiron Institute, 162 5th Ave., New York, NY 10010, USA}
\author[0000-0003-0603-8942]{Andrew Wetzel}
\affiliation{Department of Physics \& Astronomy, University of California, Davis, CA 95616, USA}
\author[0000-0003-3217-5967]{Sarah R. Loebman}
\affiliation{Department of Physics, University of California, Merced, 5200 Lake Road, Merced, CA 95343, USA}
\author[0000-0002-9933-9551]{Ivanna Escala}
\altaffiliation{Carnegie-Princeton Fellow}
\affiliation{Department of Astronomy, California Institute of Technology, 1200 E California Blvd, Pasadena, CA 91125, USA}
\affiliation{The Observatories of the Carnegie Institution for Science, 813 Santa Barbara St, Pasadena, CA 91101, USA}
\affiliation{Department of Astrophysical Sciences, Princeton University, 4 Ivy Lane, Princeton, NJ 08544, USA}
\author[0000-0002-6993-0826]{Danny Horta}
\affiliation{Astrophysics Research Institute, Brownlow Hill,  Liverpool L3 5RF}
\author[0000-0002-4900-6628]{Claude-Andr\'{e} Faucher-Gigu\`{e}re}
\affil{Department of Physics and Astronomy and CIERA, Northwestern University, 1800 Sherman Ave, Evanston, IL 60201, USA}

\correspondingauthor{Emily C. Cunningham}
\email{ecunningham@flatironinstitute.org}

\begin{abstract}
In the era of large-scale spectroscopic surveys in the Local Group (LG), we can explore using chemical abundances of halo stars to study the star formation and chemical enrichment histories of the dwarf galaxy progenitors of the Milky Way (MW) and M31 stellar halos. In this paper, we investigate using the Chemical Abundance Ratio Distributions (CARDs) of seven stellar halos from the \textit{Latte} suite of FIRE-2 simulations. We attempt to infer galaxies' assembly histories by modelling the CARDs of the stellar halos of the \textit{Latte} galaxies as a linear combination of ``template” CARDs from disrupted dwarfs, with different stellar masses $M_{\star}$ and quenching times $t_{100}$. We present a method for constructing these templates using present-day dwarf galaxies. For four of the seven \textit{Latte} halos studied in this work, we recover the mass spectrum of accreted dwarfs to a precision of  $<10\%$. For the fraction of mass accreted as a function of $t_{100}$, we find residuals of 20--30\% for five of the seven simulations. We discuss the failure modes of this method, which arise from the diversity of star formation and chemical enrichment histories dwarf galaxies can take. These failure cases can be robustly identified by the high model residuals. Though the CARDs modeling method does not successfully infer the assembly histories in these cases, the CARDs of these disrupted dwarfs contain signatures of their unusual formation histories. Our results are promising for using CARDs to learn more about the histories of the progenitors of the MW and M31 stellar halos.

\end{abstract}


\section{Introduction}

In the hierarchical paradigm of galaxy evolution, the Milky Way (MW) is more than just one galaxy. As many as thousands of dwarf galaxies are predicted to have contributed to the assembly of the MW dark matter halo over cosmic time; the remains of these accreted systems make up the component of our Galaxy known as the stellar halo. The formation conditions of halo stars are imprinted in their chemical abundances. Therefore, through observations of the chemical properties of MW halo stars, we have the unique opportunity study galaxy formation and evolution in the high redshift universe: we can study the remains of galaxies that did not survive to present day, including populations that are too faint to observe directly at high redshift, on a star by star basis. 

In the current era of MW studies, we have unprecedented knowledge of the motions and chemical properties of the stars in our Galaxy. This is in large part due to the data releases from the \textit{Gaia} mission (\citealt{GaiaCollab2016}, \citeyear{GaiaCollab2018}, \citeyear{GaiaCollab2020}) and as well as complementary wide-field spectroscopic programs, such as the GALAH survey \citep{DeSilva2015}, the H3 survey (\citealt{Conroy2019a}), APOGEE \citep{Majewski2017}, RAVE \citep{Steinmetz2006}, SEGUE \citep{Yanny2009}, and LAMOST \citep{Cui2012}. Our knowledge of the chemodynamical structure of our Galaxy will only continue to grow, with future data releases from the \textit{Gaia} mission as well as the release of data from current ongoing and planned spectroscopic surveys, including WEAVE \citep{Dalton2014}, 4MOST \citep{deJong2019}, DESI \citep{AllendePrieto2020}, and SDSS-V \citep{Kollmeier2017}. Thanks to these programs, we now have 6D phase space information as well as up to 30+ dimensions of chemical information (e.g., GALAH+ DR3, \citealt{Buder2020}) for millions of stars in the Galaxy, a truly groundbreaking achievement.

Both kinematic and chemical abundance measurements can be pertinent to a star's origin: the dynamical times of infalling and disrupting dwarfs are long compared to the age of the Galaxy (especially for stars in the outer regions of the halo). As a result, the kinematic properties of halo stars can retain a link to their initial conditions. Numerous studies have used \textit{Gaia} data plus spectroscopic information to discover new substructures in the MW halo. The early \textit{Gaia} data releases, cross-matched with spectroscopic surveys (SDSS and RAVE), enabled the discovery the ``Gaia-Sausage-Enceladus" (hereafter GSE; \citealt{Belokurov2018}, \citealt{Helmi2018}, \citealt{Koppelman2018}), a relatively metal-rich ($\langle[\rm{Fe}/\rm{H}]\rangle \sim -1.1$) population of stars on radially biased orbits, believed to be the debris from a massive, early accretion event and thought to be the dominant progenitor of the inner stellar halo. In addition, a population of stars on radial orbits, but with chemical abundances consistent with the MW thick disk, were also identified: it has been suggested that this population, referred to as the ``in-situ halo" or the ``Splash," initially formed in the MW disk but was perturbed onto halo-like orbits by early mergers (\citealt{Bonaca2017}, \citeyear{Bonaca2020}; \citealt{Haywood2018}, \citealt{DiMatteo2019}, \citealt{Belokurov2020}). While these two populations have been found to dominate local halo samples (see, e.g., \citealt{Naidu2020}), numerous additional substructures have been found through MW chemodynamical studies, including Sequoia \citep{Myeong2019}, Thamnos (\citealt{Koppelman2019}), Aleph, Arjuna, I'itoi, and Wukong \citep{Naidu2020}. Through these chemodynamical studies, many of the major building blocks of the MW have been discovered and characterized.

The remains of lower mass galaxies accreted at early times are harder to find. 
Simulations have long predicted that massive dwarfs will contribute the most by mass to the stellar halo (e.g., \citealt{Bullock2005}, \citealt{Santistevan2020}), and are even predicted to contribute the vast majority of the halo's metal poor stars (e.g., \citealt{Deason2016}). This prediction has been borne out in the observations: debris from GSE has been found to dominate nearly every observational sample of halo stars we have in the Gaia era, resulting in a stellar halo that is, on average, relatively metal-rich (\citealt{Conroy2019b}). In addition to the fact that they contribute small numbers of stars to the halo, many of the low mass dwarf galaxies accreted by the MW at early times are difficult to identify because they are likely to be phase-mixed. This is particularly challenging when studying the inner halo (where we have the highest quality spectroscopic data). While the dynamical times in the outer halo are long compared to the age of the Galaxy, enabling debris from common progenitors to be identified from kinematic measurements, the early low-mass progenitors of the inner halo are more likely be phase-mixed (e.g., \citealt{Johnston2008}), having lost their kinematic memories to the evolving MW potential. We therefore cannot rely on using kinematics alone to detect the remains of these systems: chemical abundance information is required to find stars that formed in this elusive population of galaxies.

The payoff of detecting and characterizing these low-luminosity systems within our own halo could be tremendous. For example, the faintest, early galaxies are hypothesized to be significant contributors in driving the epoch of reionization at $z>6$ (e.g., \citealt{Kuhlen2012} \citealt{Robertson2015}, \citealt{Weisz2017}, \citealt{FaucherGiguere2020}). However, even JWST will not be able to observe galaxies less luminous than the Large Magellanic Cloud directly at $z \sim 7$ (\citealt{BoylanKolchin2015}). It's long been known that archaeological studies of dwarf galaxies in the LG provide an avenue for studying faint, high-$z$ galaxies (near-field cosmology; e.g., \citealt{Freeman2002}, \citealt{Ricotti2005}, \citealt{Bovill2011}, \citealt{Brown2012}, \citealt{BoylanKolchin2015}). In principle, the disrupted dwarfs that make up the M31 and MW stellar halos can be used to increase our sample of local systems used to study faint, early galaxies with resolved stars. However, the challenge remains identifying these low-mass systems within the stellar halo. 

In this paper, we explore how the distributions of stellar chemical abundances can be used to infer the properties of stellar halo building blocks. We consider the modeling framework first proposed by \cite{Lee2015} (hereafter L15): they proposed modeling the chemical abundance ratio distribution (CARD) of a stellar halo as a linear combination of ``templates." Here, the ``templates" represent the abundance distributions of accretion events of varying stellar masses and accretion times. L15 developed and explored this method utilizing the \cite{Bullock2005} simulations, which are $N$-body simulations of accreted dwarfs onto a MW-like parent galaxy. In the context of a purely accreted stellar halo, the total halo CARD is simply a linear combination of the CARDs from all of the different accreted dwarf galaxies. Through the use of templates for different masses $M_{\rm sat}$ and accretion times $t_{acc}$, L15 posed the following question: 

``How accurately can we determine the fraction of total stellar mass, $A_j$, contributed by satellites of various mass ($M_{sat}$) and accretion time ($t_{acc}$) to a stellar halo given a set of templates for the distribution $f_j(\vec{x}_d,M_{sat},t_{acc})$ of chemical abundances $\vec{x}_d$ found in those satellites, and observations of CARDs ($f(x_d)$) in the stellar halo?"

Using this technique on the \cite{Bullock2005} simulations, L15 found that they were generally able to recover the fractional mass contributions of accreted satellites to within a factor of two, and found that this method was particularly sensitive to low mass satellites. 

L15 had the key insight that there should be information about the Galaxy's assembly history contained within the full distribution of chemical abundances in the MW halo. However, the study had its limitations in its ability to test how accurately we can constrain the assembly history from CARDs, both as a result of the chosen simulations and the available observations. In L15, because they worked with the \cite{Bullock2005} simulations, which are $N$-body simulations that include no hydrodynamics, stellar properties of the infalling dwarfs (including chemical abundances) required prescriptions. The chemical properties of the \cite{Bullock2005} halos, presented in \cite{Font2006}, were derived on an individual satellite-by-satellite basis, using the semi-analytic chemical enrichment method from \cite{Robertson2005}. While L15 showed that the assembly histories of the \cite{Bullock2005} halos could recovered reasonably well, they were not working with simulations that modeled chemical enrichment fully self-consistently. In addition, when L15 first published this technique, there were very few stars with measured kinematic and chemical properties outside of the solar neighborhood; the data available for exploiting this technique were severely limited. 

At the time of writing this publication, we are privileged to have spectroscopic metallicities measured for millions of stars in the Milky Way. Furthermore, deeper spectroscopic observations of M31 have enabled $\alpha-$abundance measurements and iron metallicities derived from full spectral synthesis for an ever growing sample of individual stars in M31's halo (\citealt{Vargas2014b}, \citealt{Escala2019}, \citeyear{Escala2020a}, \citeyear{Escala2020b}, \citeyear{Escala2021}; \citealt{Gilbert2019}, \citeyear{Gilbert2020}) and M31's satellites (\citealt{Vargas2014a}, \citealt{Kirby2020}, \citealt{Wojno2020}). With more knowledge than ever about the chemodynamical structure of the Local Group, we are now in a position to explore more deeply the potential of utilizing CARDs to constrain the formation histories of Local Group galaxies, and address how this method might aid in identifying stars that were born in high-redshift low-mass galaxies that cannot be observed directly. 

In this paper, we undertake a critical next step in exploring the feasibility of using CARDs to constrain the a galaxy's assembly history: we explore how CARDs can be leveraged to constrain the formation histories of the FIRE-2 zoom-in cosmological simulations of Milky Way-mass galaxies (introduced in \citealt{Wetzel2016}). These simulations self-consistently model chemical enrichment, and contain many complications that are not included in the \cite{Bullock2005} models. Examples of such complications include dwarf-dwarf interactions, live dark matter halos and disks, the time variability of the host potential, and the role of feedback in shaping dwarf galaxy star formation histories. 

In order to make the CARDs method into a practical method to apply to observations, we seek to address three key questions in this work:
\begin{enumerate}
    \item Does the method proposed by L15 of modeling the stellar halo CARD as a linear combination of templates work in a more realistic setting?
    \item We require a strategy for an observationally viable method for constructing these templates. Can we use the CARDs of present-day dwarf galaxies to infer the properties of the disrupted dwarfs?
    \item How can we assess the accuracy of this method? 
\end{enumerate}

It is crucial that we definitively answer this second question, as observationally, the construction of empirical templates for halo progenitors are likely to be limited to a sample of nearby present-day dwarf galaxies, where member stars can be identified with high confidence. Abundance distributions have been measured already for a number of local dwarf galaxies (see, e.g., \citealt{Kirby2009}, \citeyear{Kirby2010}, \citeyear{Kirby2017}, \citeyear{Kirby2020}, \citealt{Nidever2020}, \citealt{Hasselquist2021}). The number of satellites with abundances measured over their full spatial extent, and to fainter magnitudes, will only increase with upcoming surveys; for example, mapping these distributions is one of the main objectives of the Subaru Prime Focus Spectrograph Galactic Archaeology program \citep{Takada2014}. 
Therefore, from an observational perspective, local dwarf galaxies would be the ideal choice as the basis for empirical template CARDs for stellar halo building blocks. 

However, it is well known that MW halo stars have higher $\alpha-$abundances relative to their dwarf galaxy counterparts (e.g., \citealt{Venn2004}). This is a natural consequence of the fact that halo stars formed in dwarfs that did not survive until present-day (e.g., \citealt{Robertson2005}): dwarf galaxies that survive until present-day are more likely to have extended star formation histories than halo progenitors. Therefore, if using stars in dwarf galaxies to constrain the properties of halo progenitors, the potentially different timescales in star formation must be taken into account. In addition, given that present-day dwarfs are likely to be in more isolated environments at early times than halo progenitors, these different environments could have an effect on their chemical enrichment and evolution (e.g., \citealt{Corlies2013}). 

In this work, we demonstrate a method for creating templates for the progenitors of stellar halos using exclusively present-day dwarf galaxies (accounting for the fact that the two populations are likely to be forming stars on different timescales), and assess the performance of these templates relative to the templates constructed from streams and phase-mixed debris. While an in-depth exploration and comparison of the chemical enrichment histories of present-day dwarfs versus halo progenitors is beyond the scope of this work, we can investigate whether halo progenitor CARD templates can be reliably constructed from a sample of dwarf galaxies. 

This paper is organized as follows. In Section \ref{sec:sims}, we introduce the simulations used in this work. We summarize the key relevant details about the FIRE-2 zoom-in cosmological simulations in Section \ref{sec:fire_sims} and introduce the catalog of halo progenitors and present-day dwarf galaxies in Section \ref{sec:p21_cat}. In Section \ref{sec:context}, we explore the CARDs in the simulation and demonstrate their connection to the formation history of the host halo (in Section \ref{sec:dform_age}) and the dynamical state of the accreted satellite (i.e., classification as phase-mixed debris, coherent stream or dwarf galaxy; in Section \ref{sec:dyn_cards}). In Section \ref{sec:methods}, we describe the CARDs technique (in Section \ref{sec:model}) and our method for constructing CARD templates (in Section \ref{subsec:temps}). In Section \ref{sec:results}, we present the results of our modeling procedure for all the simulations used in this work, and discuss in detail a successful case as well as a less successful case. 
In Section \ref{sec:discussion}, we discuss some of the limitations of this technique, and discuss potential improvements and extensions of the method presented here.
We conclude in Section \ref{sec:concl}.

\begin{figure*}
    \centering
    \includegraphics[width=\textwidth]{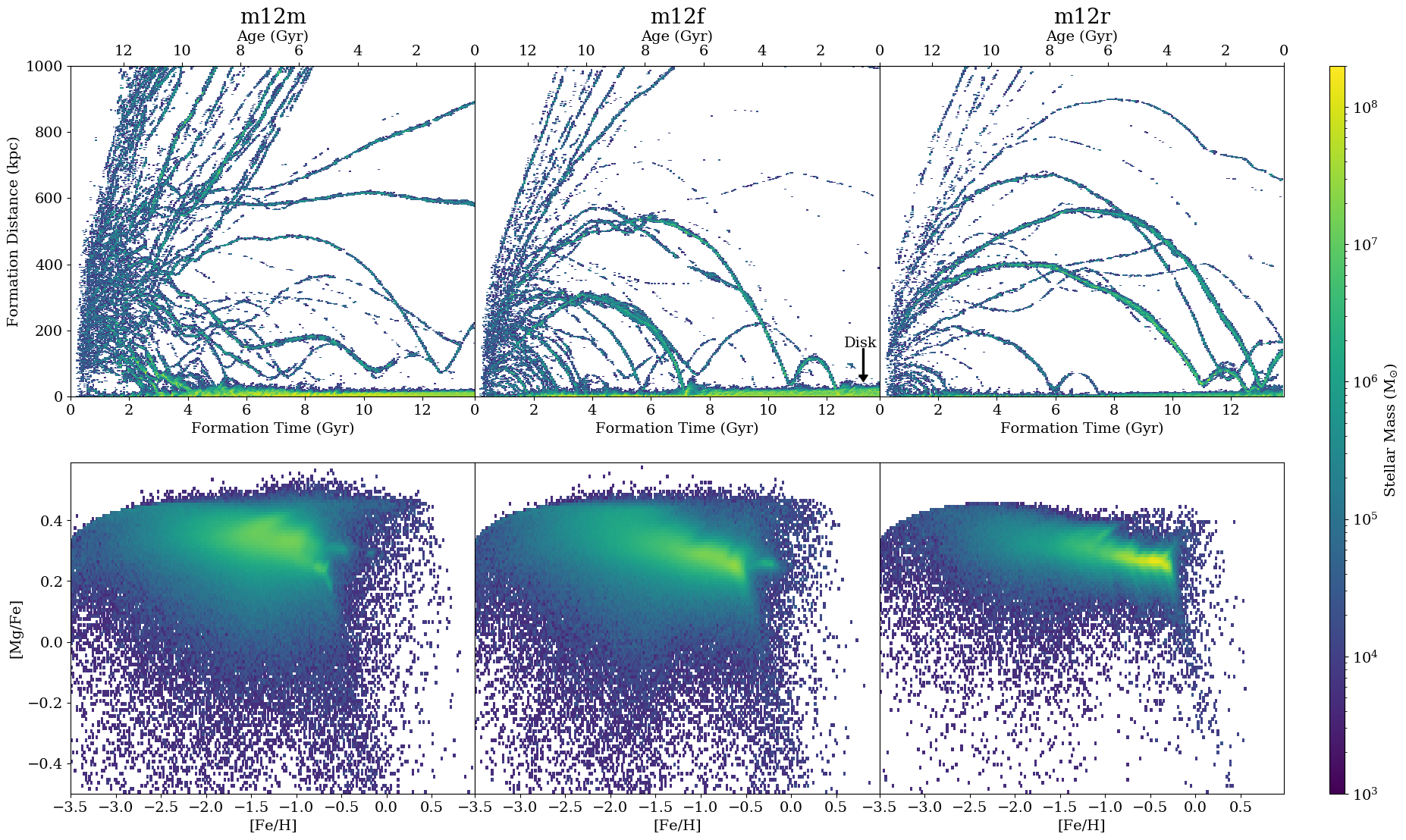}
    \caption{Top panels: Formation histories for three of the \textit{Latte} simulations (m12m, m12f and m12r), as shown by 2D histograms of formation distance versus formation time (defined such that $t_{\rm form}=0$ at the beginning of the simulation), for all star particles. Density is normalized on a logarithmic scale. In this plane, the thick regions of high density at small values of formation distance show star particles forming in the disks of the main host galaxies. The lower density streaks that start at larger formation distance and move towards lower formation distances indicate star particles forming in dwarf galaxies and falling into the host. From these figures, we see that m12m's assembly history (left) is characterized by early accretion, though it also has several massive dwarfs in the halo at present day. In contrast, m12r (right) experiences several recent, massive accretion events, with less activity at early times. The intermediate case is shown by m12f (center), which experiences a relatively recent massive accretion ($\sim 2$ Gyr ago) event as well as one $\sim 6$ Gyr ago. Lower panels: the corresponding CARD for the stellar halo of each for galaxy, taking star particles that have formation distances greater than 30 kpc from the main host, but that lie within 100 kpc of the main host at present day. These visibly different distributions reflect the formation histories of each of their hosts. For example, m12m (left), characterized by early accretion, has higher density at lower \feh and higher [Mg/Fe]. In contrast, the recent accretion experienced by m12r (left) results in an abundance distribution that has high density at high \feh and lower [Mg/Fe].}
    \label{fig:form_age}
\end{figure*}

\section{Simulations}
\label{sec:sims}
In order to test the CARDs method in a more realistic setting, in this work,  we make use of two suites of cosmological zoom-in simulations from the Feedback in Realistic Environments (FIRE) project.\footnote{FIRE project website: \url{http://fire.northwestern.edu}}. We begin in Section \ref{sec:fire_sims} by summarizing the relevant properties of the simulations; in Section \ref{sec:p21_cat}, we introduce the subset of stellar halo and dwarf galaxy star particles from the simulation that we use in this analysis.

\subsection{FIRE-2 Simulations}
\label{sec:fire_sims}
All of the simulations used in this work were run using the GIZMO\footnote{\url{http://www.tapir.caltech.edu/~phopkins/Site/GIZMO.html}} gravity plus hydrodynamics code in meshless finite-mass (MFM) mode (\citealt{Hopkins2015_gizmo}) with the FIRE-2 physics model (\citealt{Hopkins2018_fire2}). While we refer the reader to the above papers for details of the FIRE-2 implementation, we summarize some of the key details here, especially those related to star formation and chemical enrichment. 

The FIRE-2 gas model includes metallicity-dependent treatment of radiative heating and cooling processes, over a temperature range of $10-10^{10}$ K (\citealt{Hopkins2018_fire2}). The FIRE-2 simulations were evolved with the \cite{FaucherGiguere2009} UV background model, in which reionization occurs early ($z_{\rm reion}\sim10$). Star particles form from self-gravitating, cold, dense, molecular, Jeans-unstable gas clouds (following \citealt{Krumholz2011}). Star particles are treated as individual stellar populations with a \cite{Kroupa2002} initial mass function. Type II SNe rates, as well as mass and metal loss rates, are derived from stellar population models (STARBURST99; \citealt{Leitherer1999}). Metal yields for Type Ia SNe are taken from \cite{Iwamoto1999} and yields for Type II SNe are from \cite{Nomoto2006}. 

The three sources of chemical enrichment in the FIRE-2 simulations are Type Ia SNe, Type II SNe, and stellar winds (produced predominantly by asymptotic giant branch stars and O-type stars). The simulations self-consistently track eleven chemical abundances: H, He, C, N, O, Ne, Mg, Si, S, Ca, and Fe. We utilize the runs implementing sub-grid turbulent metal diffusion (\citealt{Hopkins2017_diffusion}), which \cite{Escala2018} demonstrated result in realistic spreads in the abundance distributions in dwarf galaxies, while not resulting in significant changes to star formation (\citealt{Su2017}). 

In this work, our primary focus is a set of seven galaxies from the \textit{Latte} suite of FIRE-2 simulations (introduced in \citealt{Wetzel2016}). The \textit{Latte} suite are simulations of isolated galaxies with present-day masses in the range of $M_{200m}=1-2 \times 10^{12}~M_{\odot}$, similar to the MW and M31. These simulations are high resolution, with initial masses of star particles with $\sim 7000 M_{\odot}$. While the $z=0$ particle masses depend on age, the average particle mass is $\sim 5000~M_{\odot}$, as a result of stellar mass loss.

While we focus on the stellar halos of seven of the isolated MW analogs from the \textit{Latte} suite in this paper, we also make use of the satellites from the ELVIS on FIRE suite (\citealt{GarrisonKimmel2019a}, \citealt{GarrisonKimmel2019b}) to increase our sample of dwarf galaxies. The ELVIS on FIRE suite consists of three simulations, each containing a M31--MW analog pair; the DM halos selected for zoom-in simulation were chosen to have the approximate distance and relative velocities of the MW and M31 (\citealt{GarrisonKimmel2019a}). 
While the physics in the \textit{Latte} runs and the ELVIS on FIRE runs we use in this work are the same, we note that the mass particle resolution in the ELVIS on FIRE suite is approximately twice as high, with initial star particle masses of $\sim 3500-4000~M_{\odot}$; however, for the analysis presented here, the difference in resolution does not affect the outcome.

The properties of the host galaxies in these simulations have been demonstrated to show broad agreement with the MW and M31, including the stellar-to-halo mass relation (\citealt{Hopkins2018_fire2}), stellar halos (\citealt{Sanderson2018}, \citealt{Bonaca2017}), and the radial and vertical structure of their disks (\citealt{Ma2017}, \citealt{Sanderson2020}, \citealt{Bellardini2021}). 
Critically for this work, the satellite populations of both of these suites of simulations have also been shown to agree with many observed properties, such as their masses and velocity dispersions (\citealt{Wetzel2016}; \citealt{GarrisonKimmel2019a}), star formation histories (\citealt{GarrisonKimmel2019b}), and radial distributions (\citealt{Samuel2020a}). 

While the satellite galaxies in these simulations have been shown to have many properties in common with the observed satellites around the MW and M31, they are generally found to be too metal-poor compared to the observed dwarf galaxies (\citealt{Escala2018}, \citealt{Wheeler2019}, \citealt{Panithanpaisal2021}). This is likely at least in part because of the assumed delay-time distributions for SNe Ia. These simulations adopt a SNe Ia rate with prompt and delayed components (\citealt{Mannucci2006}); adopting a SNe Ia delay time distribution that results in a larger total number of SNe Ia events (e.g., a power-law rate; \citealt{Maoz2017}) helps to resolve the discrepancy between the simulated and observed iron abundances (Gandhi et al., in prep). For this work, we emphasize that we do not require quantitative agreement with simulated and observed abundances. \textbf{The empirical CARD templates we construct in this work from the simulations are intended for use within the simulations only.} While the method presented here of using empirical templates to infer halo progenitor properties can be applied to observational data, we emphasize that the template CARDs from the simulations should not be applied to the observations. 

\subsection{Halo Progenitors and Dwarf Galaxies}
\label{sec:p21_cat}

In investigating how CARDs of stellar halos can be used to infer assembly histories, and in comparing the abundance distributions of disrupted dwarfs with the present-day dwarfs, we must first define our stellar halo and dwarf galaxy samples within the simulation. In this section, we introduce the catalog of halo star particles and dwarf galaxies that we use for this experiment. We limit our study to the catalog of star particles belonging to dwarf galaxies, stellar streams, and phase-mixed debris constructed in \cite{Panithanpaisal2021} (hereafter P21)\footnote{Stream catalog publicly available: \url{https://flathub.flatironinstitute.org/sapfire}}. 
By using this subset of particles, we can explore the efficacy of the CARDs framework for a purely accreted stellar halo in a more realistic setting than L15 while retaining control of the precise composition of the accreted stellar halo, and make comparisons between halo progenitor and present-day dwarf galaxy populations. We explore these questions in the limit of no contamination from disk or in-situ halo stars, and leave strategies for modeling potential contamination to future work. 

While the primary goal of P21 was to identify present-day streams, as a result of their stream candidate identification method, they also found found present-day dwarf galaxies and phase-mixed debris. P21 implemented the following criteria to identify stream candidates:

\begin{itemize}
    \item Particles in stream candidates must be within the virial radius of the main host at present day, though bound in a different subhalo 2.7--6.5 Gyr ago.
    \item Stream candidates must have $120<N<10^{5}$ star particles; this results in a stellar mass limit for the progenitors of $10^{5.5}M_{\odot}<M_{\star}<10^{9}M_{\odot}$ for the isolated MW simulations.\footnote{This corresponds to an upper mass limit of $\sim M_{\star}<10^{8.5} M_{\odot}$ for the paired simulations, which have typical particle masses of $\sim 3000 M_{\odot}$.}
\end{itemize}

With stream candidates in hand, each candidate is classified as either phase-mixed debris, a coherent stream, or a dwarf galaxy based on its local velocity dispersion and maximum pairwise distance between star particles at present day. To be classified as a coherent stream or as phase-mixed debris, star particles must have a maximum pairwise distance greater than 120 kpc. This pairwise distance threshold, on the order of the size of the host galaxy halo, effectively distinguishes between disrupted systems (phase-mixed debris and coherent streams) and present-day bound dwarf galaxies, which remain compact in position space. P21 used a linear kernel Support Vector Machine (SVM) in order to derive an average local velocity dispersion threshold as a function of stellar mass to distinguish phase-mixed debris from coherent streams (see their Equation 2). Using disrupted systems from four of the simulations, P21 classified coherent streams and phase-mixed debris by eye to use as the training set for the SVM. In practice, coherent streams have lower average local velocity dispersions than phase mixed debris (at fixed $M_{\star}$). P21 determines the local velocity dispersion for a star particle using its nearest neighbors in phase space (using 20 nearest neighbors if the stream candidate has more than 300 star particles; otherwise, 7 nearest neighbors are used); they then average over all star particles in the stream candidate to get the average local velocity dispersion. This local velocity dispersion threshold is $\sim 20$ \kms ~for a stream candidate with \mstar $\sim 10^7$ \msun ~(see their Figure 1). 

In summary, present day dwarf galaxies are compact in velocity and position space; coherent streams have low local velocity dispersions but are extended in position-space; and phase-mixed debris have larger local velocity dispersions and can be extended position space. We define the stellar halo samples analyzed in this work to include the phase-mixed debris and streams from P21, and exclude the present-day dwarf galaxies from our stellar halo samples.

We note that not every accreted star particle in the simulation is included in the P21 catalog. In particular, accreted dwarfs that exceed the particle limit from P21 (with \mstar $~>10^9$ \msun, bordering on major merger classification) are not included, as well as low-mass systems containing fewer than 120 star particles ($\sim 10^5 M_{\odot}$). In addition, a number of early mergers (that are unbound prior to 6.5 Gyr ago) are also excluded (these earlier mergers are discussed in more detail in Horta et al., in prep). Table \ref{tab:exsitu} summarizes the total fraction of the accreted stellar halo that is included in the P21 catalog for the seven halos analyzed in this work. To estimate the total mass of the accreted stellar component, we compute the mass of all star particles that are located within the virial radius of the main host at present day, but that formed at least 30 kpc from the main host (excluding stars bound in satellites). The total accreted stellar masses range from $1-11\times10^9 M_{\odot}$. Table \ref{tab:exsitu} shows that the fraction of accreted star particles included in the P21 varies widely from simulation to simulation. Halos that have a relatively large accreted mass usually have experienced at least one massive accretion event above the P21 limit; the two halos with the largest accreted stellar masses (m12b and m12r) have the lowest fraction of their accreted stars included in the P21 catalog. For all simulations, we can see that material that is accreted early does contribute significantly to the stellar mass of the overall halo; these accretion events are explored in more detail in Horta et al. (2021, in prep).   

While there are many accreted halo star particles that are not included in our analysis, we note that we do not require a fully complete halo sample to test the validity of this method. In order to address whether or not the CARDs method works in a cosmological simulation, we require a sample of accretion events whose progenitors have been well characterized. We therefore limit the scope of this study to using accretion events and dwarf galaxies from the P21 catalog.

\begin{table}[]
    \centering
    \begin{tabular}{c|c c}
    Simulation & $M_{\star, \mathrm{acc}}$ ($10^9 M_{\odot}$) &  \% Stellar Halo Mass in P21 \\
    \hline \hline
      m12b   & 11.1 & 2.4 \\
      m12c   & 1.8 & 26.4 \\
      m12f   &  6.6 & 9.3 \\
      m12i   & 2.9 & 18.0 \\
      m12m   &  3.8 & 9.3 \\
      m12r   & 7.8 & 0.4 \\
      m12w   & 2.9 & 14.6 \\
    \end{tabular}
    \caption{Each of the simulations discussed in this work, with the fraction of accreted stars (excluding those bound in satellites) included in the P21 catalog. To compute the total accreted contribution, we calculate the total mass in star particles that are located within the virial radius at present day, but that formed at a distance larger than 30 kpc from the center of main halo. We exclude star particles in bound subhalos at present day from this estimate. Halos with the largest total accreted stellar masses have the lowest fraction of their halos included in P21, as these halos experience at least one major accretion event above the mass limit imposed by P21.}
    \label{tab:exsitu}
\end{table}

\begin{figure*}
    \centering
    \includegraphics[width=0.24\textwidth]{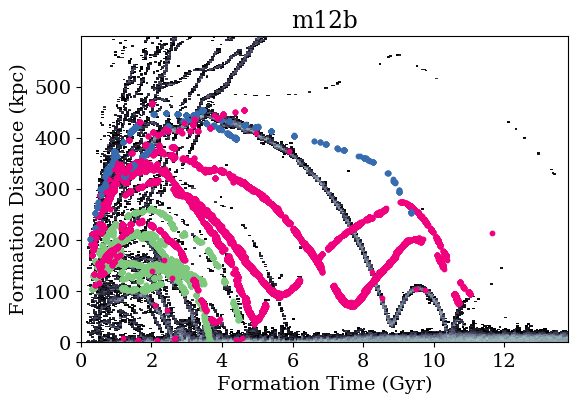}
     \includegraphics[width=0.24\textwidth]{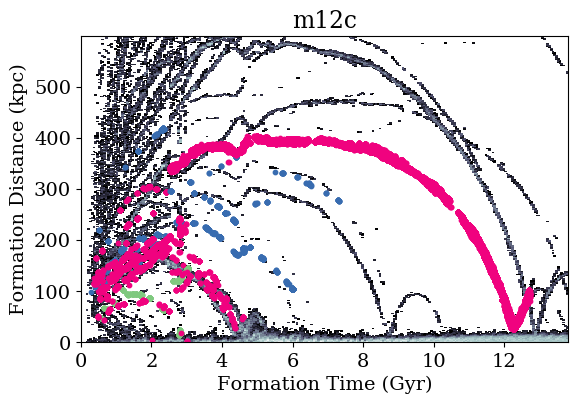}
      \includegraphics[width=0.24\textwidth]{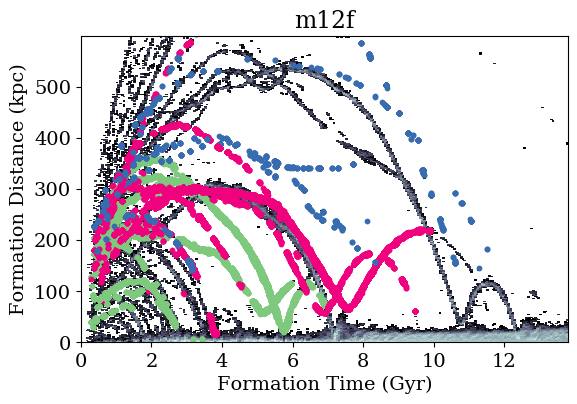}
       \includegraphics[width=0.24\textwidth]{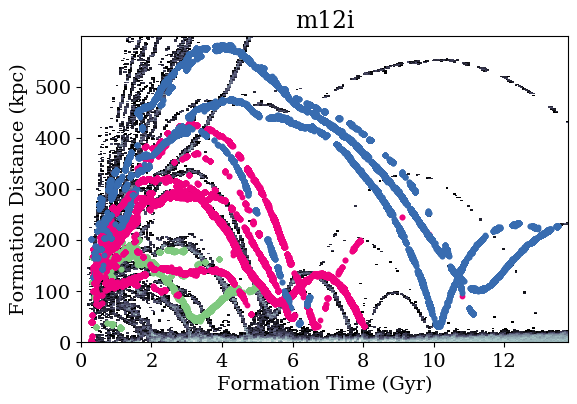}
        \includegraphics[width=0.24\textwidth]{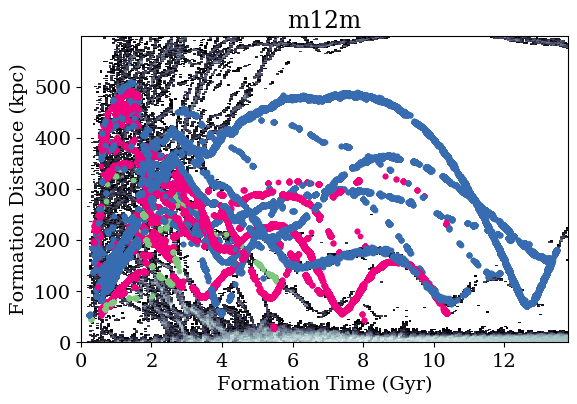}
         \includegraphics[width=0.24\textwidth]{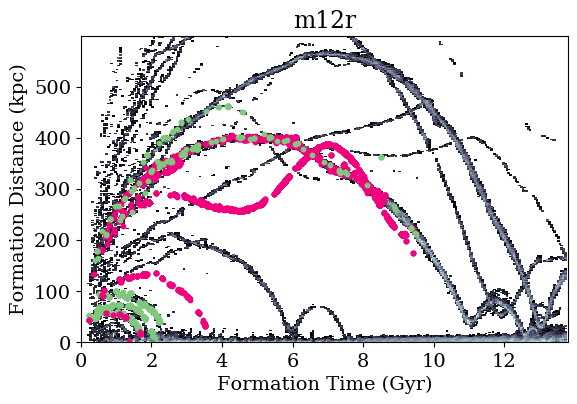}
          \includegraphics[width=0.24\textwidth]{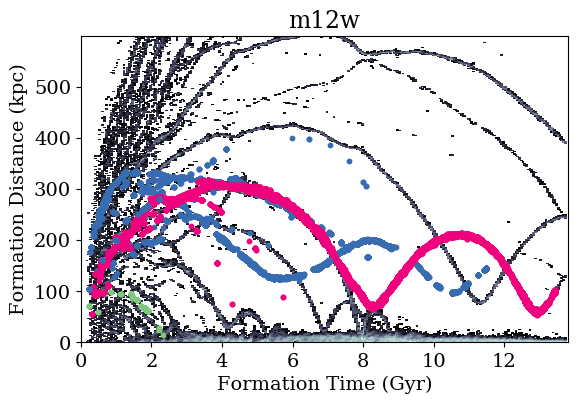}
          \includegraphics[width=0.24\textwidth]{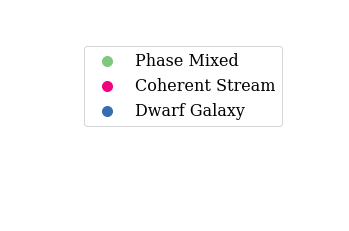}
    \caption{Same as the top panels of Figure \ref{fig:form_age}, but for the set of seven \textit{Latte} simulations studied in this work. Star particles from the P21 stream candidates catalog (used in this work) are indicated by the colored points. Star particles belonging to surviving dwarf galaxies (within virial radius of the host galaxy at present day in the simulations) are shown in blue; particles classified as members of coherent stellar streams are in pink; and phase-mixed debris are shown in green. Not every star particle that is accreted in the simulation is included in the P21 catalog (and therefore in this analysis); see Section \ref{sec:p21_cat} for details.}
    \label{fig:form_age_acc_parts}
\end{figure*}

\section{Simulation CARDs}
\label{sec:context}

We seek to use the chemical abundance distribution of halo stars to infer properties of their dwarf galaxy progenitors. In this section, we demonstrate how halos with different formation histories come to have different CARDs (Section \ref{sec:dform_age}), in order to motivate why we can use halo star CARDs to make inferences about dwarf galaxy progenitors of stellar halos. In Section \ref{sec:dyn_cards}, we demonstrate the relationship between a halo progenitor CARD and its dynamical state (e.g., whether or not they are phase-mixed, coherent streams or present-day dwarfs). 

\subsection{The Link between CARDs and Assembly History}
\label{sec:dform_age}

In this subsection, we demonstrate the relationship between halo CARDs and host halo formation history, by discussing in detail three hosts from the \textit{Latte} suite of simulations. The top panels of Figure \ref{fig:form_age} show 2D histograms of formation time versus formation distance for every star particle in simulations m12m, m12f, and m12r (with density plotted on a logarithmic scale). We note that formation time is defined such that $t_{\rm form}=0$ at the beginning of the simulation, and the formation distance is physical distance (as opposed to comoving distance). Looking in this plane, we can see where the star particles are forming in these different simulated galaxies over time. The high-density regions at the bottom of each figure, showing star particles with formation distances $0<d_{\rm form}<20$ kpc, are particles that form in the disk of the host galaxy. 
The streaks of star particles that start at larger $d_{\rm form}$ and move towards the disk over time show dwarf galaxies falling into the main host halo. Figure \ref{fig:form_age} demonstrates that these three simulations have very different formation histories: for example, m12m's evolutionary history (shown in the top left panel of Figure \ref{fig:form_age}) is characterized by many early mergers, whereas m12r (top right panel of Figure \ref{fig:form_age}) experiences two massive, recent accretion events ($M_{\star}>10^9 M_{\odot}$, within the last $\sim 2$ Gyr). We note that m12r was one of the \textit{Latte} hosts chosen with the intention of simulating a MW host with an LMC-analog; see \cite{Samuel2020a}.

The lower panels of Figure 1 show the chemical abundance ratio distributions (CARDs) for these same galaxies, in [Mg/Fe] versus \feh; abundances are shown for all stars that formed outside of 30 kpc but are within 100 kpc of the host galaxy at present day. These CARDs are visibly quite different, with apparent links to the formation histories of their host halos. 

The galaxy m12m (left panels of Figure \ref{fig:form_age}) experiences a number of massive accretion events early on, and while there are many dwarfs in the vicinity of m12m at present day, most have large pericenters (as seen by the fact that the wiggles do not intersect with the disk). Most of the density in the CARD for m12m is at lower \feh ($\sim-2<$\feh$-1$) and higher [Mg/Fe] ([Mg/Fe]$>0.25$). In contrast, m12r (right panels of Figure \ref{fig:form_age}) has an assembly history characterized by recent, massive accretion events, indicated in the top panel of Figure \ref{fig:form_age} by the thick streaks of high density that approach the disk at young ages/recent times. The resulting CARD shows high density at higher \feh and low [Mg/Fe] relative to m12m, consistent with the picture that the halo is dominated by dwarfs that were accreted recently and had a prolonged star formation and chemical enrichment history. Finally, m12f has experienced several massive accretion events over the course of its formation, with a major event recently ($\sim 2$ Gyr ago) and another major event at intermediate times ($\sim 6$ Gyr ago); the resulting CARD shows density at both low \feh and high [Mg/Fe] as well as higher \feh and lower [Mg/Fe].

These different distributions arise due to the different timescales of chemical enrichment. At early times, galaxies are enriched by Type II supernovae (beginning around $\sim 4$ Myr in the simulation), and producing iron and the $\alpha-$elements at a relatively constant fraction. As a result, early in a galaxy's lifetime, the mean \afe is roughly constant. Once the Type Ia supernovae turn on (at first around 40 Myr, with the rate dependent on the star particle age), the \afe ratio of stars born in the enriched ISM starts to decrease, because Type Ia supernovae produce more iron than they do $\alpha-$elements. 
Therefore, the CARD of the stellar halo in a galaxy contains key information about the star formation histories of the halo progenitors.

However, what remains an open question is whether or not we can use CARDs from present-day dwarf galaxies to infer the properties of halo progenitors. In the next section, we introduce the chemical properties for the catalog of dwarf galaxies and halo progenitors used in this work. 

\begin{figure*}
    \centering
    \includegraphics[width=\textwidth]{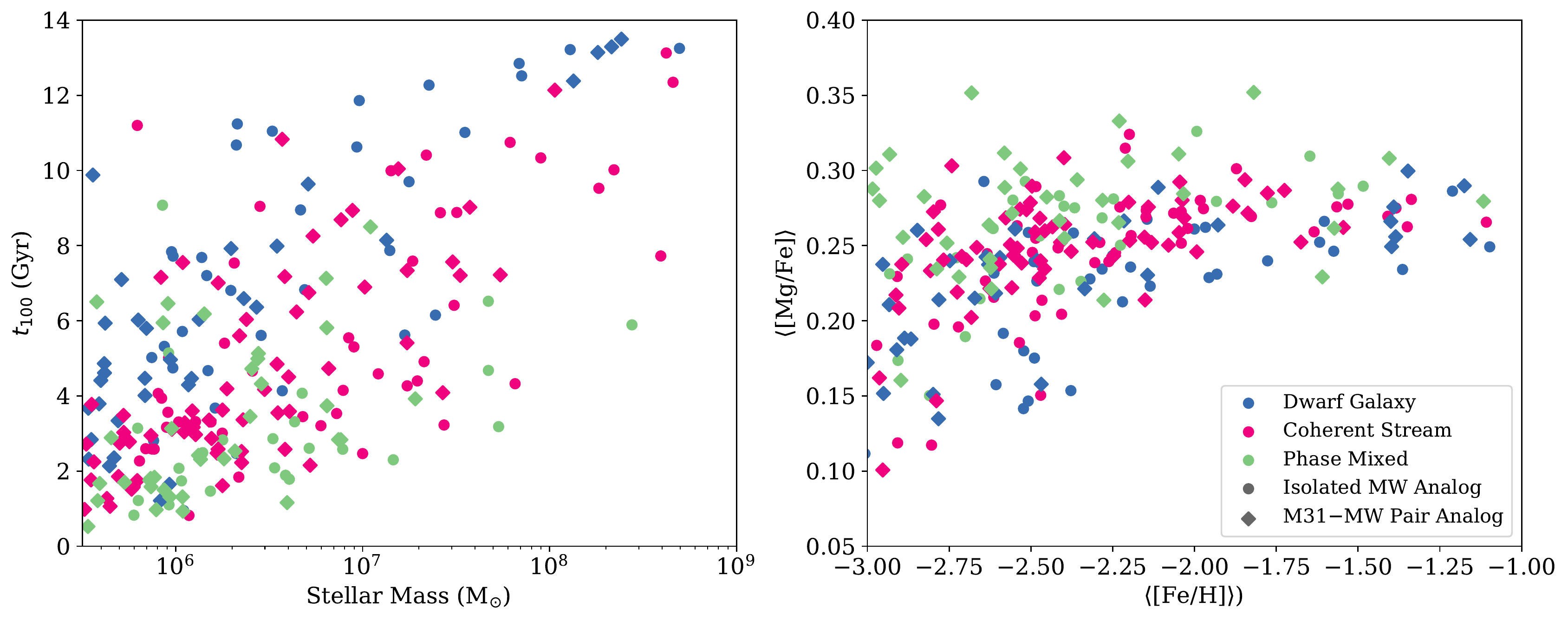}
    \caption{Left: stellar mass vs quenching time ($t_{100}$) for the full P21 catalog. Points are color coded based on their classification as dwarf galaxies, coherent streams, or phase mixed debris. Diamond shaped points are for objects in the paired M31--MW analog simulations; circular points are objects from the isolated MW analog simulations. While there is substantial overlap in this plane across the three classifications, at a given stellar mass, dwarf galaxies tend to have the latest quenching times and phase-mixed debris tends to have the earliest quenching times. Right: mean [Mg/Fe] vs [Fe/H] for the full P21 catalog. At fixed metallicity, phase-mixed debris is, on average, more enhanced in Mg than dwarf galaxies.}
    \label{fig:mstar_deltat}
\end{figure*}

\subsection{Linking Dynamical State and Chemical Abundances}
\label{sec:dyn_cards}

Chemical abundance distributions can be used to ``tell time" within a dwarf galaxy or halo progenitor: because the different chemical enrichment processes (e.g., Type Ia SNe, Type II SNe, and stellar winds) occur on different timescales, the distribution of a dwarf or progenitor's chemical abundances is informative for determining how much stellar mass formed when in the system's evolution. An additional clue constraining the amount of time a halo progenitor had to form stars is its \textit{dynamical state} at present day, which we take here to mean whether or not a halo progenitor is phase-mixed, a coherent stream, or remains a dwarf galaxy at the end of the simulation. Because these different classifications imply different times for truncating star formation (with phase-mixed debris, on average, implying an earlier accretion time than a surviving dwarf), their chemical properties are in turn expected to be different. In this section, we compare the chemical properties of phase-mixed debris, coherent streams and dwarf galaxies from the P21 catalog. This has implications for using the CARDs technique in an observational context, where we will likely need to rely on a sample of present-day dwarf galaxies for constructing an empirical template set to model the MW's halo progenitors. 

Figure \ref{fig:form_age_acc_parts} shows the 2D histogram of age versus formation distance for the seven of the isolated MW simulations used in this work in gray, with the star particles included in the P21 catalog overplotted in color. Points are color-coded by classification: dwarf galaxies are in blue, coherent streams are in pink, and phase-mixed debris are in green. 

Figure \ref{fig:form_age_acc_parts} highlights some systematic differences between the dwarf galaxy, stream and phase-mixed populations. For example, most of the star particles associated with phase-mixed debris (colored in green) are primarily composed of older star particles, consistent with the picture that they must have been accreted relatively early on in order to have had sufficient time to become phase-mixed at present day. The differences between the populations are highlighted in more detail in Figure \ref{fig:mstar_deltat}. The lefthand panel of Figure \ref{fig:mstar_deltat} shows $t_{100}$ versus total stellar masses for all objects from the P21 catalog, where we define $t_{100}$ to be the time for a dwarf or halo progenitor to form 100\% of its stars (i.e., the quenching time). As previously, points are color-coded by classification (dwarf galaxies in blue, coherent streams in pink, and phase-mixed objects in green) while the marker shape indicates whether or not the object is in an isolated MW simulation (circles) or in a paired galaxy simulation (diamonds; the ELVIS on FIRE suite). While there is substantial overlap in this plane amongst the three classifications, at fixed stellar mass, phase mixed objects have smaller $t_{100}$ values (on average) than the surviving dwarf galaxies, as a result of the fact that they are disrupted earlier and have less time to form more stellar mass. This has implications for the abundance distributions (which we can directly observe), as shown by the right panel of Figure \ref{fig:mstar_deltat}: at fixed mean metallicity ($\langle [\mathrm{Fe/H}] \rangle$), phase-mixed debris have higher mean magnesium abundances compared to the surviving dwarf galaxies. We note that while there is a correlation between mass and metallicity in the simulations (e.g., \citealt{Wetzel2016}, \citealt{Escala2018}, P21), there is not a direct correlation between mean [Mg/Fe] and $t_{100}$. As the left panel highlights, $t_{100}$ is also correlated with stellar mass, and therefore mean [Fe/H]. Therefore, the quenching time $t_{100}$ is only correlated with mean [Mg/Fe] at fixed stellar mass.

The fact that the phase-mixed debris is, on average, $\alpha-$enhanced (at fixed metallicity) with respect to present-day dwarf galaxies is consistent with a long-known observational result: halo stars are observed, on average, to be $\alpha$-enhanced relative to stars that formed in Local Group dwarf galaxies (see, e.g., \citealt{Venn2004}). Figure \ref{fig:mstar_deltat} demonstrates that this arises because halo stars form in dwarfs that disrupt before $z=0$, and do not not have as much time to enrich their ISMs in iron through Type Ia SNe compared to present-day dwarfs. Comparing present-day dwarf populations with halo star populations therefore results in a ``temporal bias" (e.g., \citealt{Robertson2005}, \citealt{Font2006}, \citealt{Johnston2008}): at fixed stellar mass, dwarf galaxies have, on average, spent a longer period of time forming their stars. Therefore, if we are to use dwarf galaxies to recover the properties of accreted debris, we must correct for this effect in constructing our templates. In the subsequent section, we describe our method for using the CARDs for present-day dwarf galaxies as templates for halo progenitors. 

\begin{figure*}
    \centering
    \includegraphics[width=\textwidth]{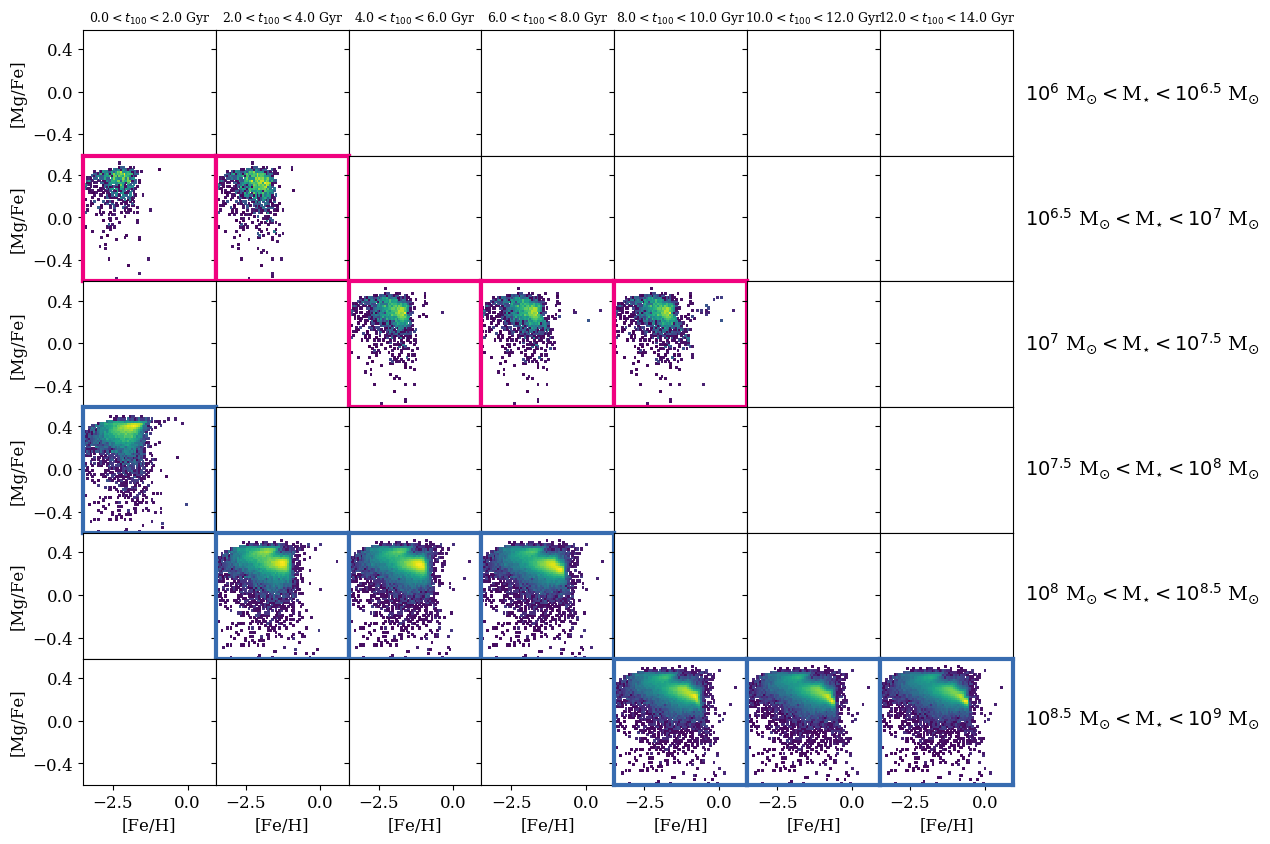}
    \caption{A grid of templates, including only two sources from the P21 catalog in the m12m simulation, to introduce our grid axes and demonstrate our procedure for creating templates. All of the templates shown in the bottom three rows (outlined in blue) are created from a single present-day dwarf galaxy, with $M_{\star}=10^{8.7} M_{\odot}, t_{100, \mathrm{source}}=13.25$ Gyr. To create the templates with $0<t_{100, \mathrm{temp}}<2$ Gyr, we consider only the CARDs for the accretion events for star particles that formed within the first 2 Gyr of the simulation. Because the massive dwarf is still forming stars at present day in the simulation, it contributes to each bin of $t_{100, \mathrm{temp}}$; it moves to progressively larger mass bins as it forms stars throughout the simulation. The templates shown in the second and third rows (outlined in pink) are created from a present-day coherent stream ($M_{\star}=10^{7.15} M_{\odot}, t_{100, \mathrm{source}}=10.0$ Gyr). Because the stream stops forming stars after 10 Gyr, it does not contribute to the templates in the far two left columns (where $t_{100, \mathrm{temp}}>t_{100, \mathrm{source}}$). }
    \label{fig:temp_demo}
\end{figure*}

\section{Methodology}
\label{sec:methods}

With the CARDs method, first proposed by L15, we model the abundance distributions in the stellar halo as a linear combination of template CARDs. Two of the primary goals of this paper are to test the CARDs modeling procedure in the more realistic setting of a cosmological simulation, and to explore how to construct template CARDs in a potentially observationally viable manner. In this section, we describe the CARDs modeling method, and present our method for constructing empirical templates.

We begin by describing the CARDs modeling method in Section \ref{sec:model}; for the reader's convenience, we have summarized a number of relevant definitions and notation in Table \ref{tab:defn}. In Section \ref{subsec:temps}, we describe our procedure for creating empirical templates, and introduce the template sets used for analysis in this work. 

\subsection{Modeling Framework}
\label{sec:model}

In this subsection, we demonstrate our procedure for modeling the CARDs of the \textit{Latte} halos using a grid of template abundance distributions. For the purposes of keeping our notation general, we define the vector of chemical abundance ratios as $\vec{x}_d$: in this work, we limit ourselves to the cases where $\vec{x}_d=\left( \mathrm{[Fe/H], [Mg/Fe]}\right)$ or $\vec{x}_d=\left( \mathrm{[Fe/H], [Mg/C]} \right)$ (though, in  principle, this method could make use of many more elements and dimensions). While the simulations track 11 elements, we use the elements most dominated by each of the three channels of chemical enrichment (Fe for Type Ia SNe, Mg for Type II SNe, and C for stellar winds).
We define our grid of templates over a range of stellar mass and quenching time. Bins in stellar mass are denoted with the index $i$ and bins in quenching time are denoted with the index $j$ (we define the specifics of the grid used in this work in Section \ref{subsubsec:temp_axes}). 

As in L15, we express the model for the (normalized) full abundance distribution of the halo (which we denote as CARD$_{\rm halo}$) as a linear combination of individual templates:

\begin{multline}
   \mathrm{CARD}_{\rm halo, model} (\vec{x}_d) = \\ \sum_{i} \sum_{j} A_{ij} \mathrm{CARD}_{\mathrm{temp}, ij}(\vec{x}_d | M_{sat,i}, t_{100,j}),
   \end{multline}
where $A_{ij}$ is the coefficient for the template for an accretion event with stellar mass $M_{sat,i}$ and quenching time $t_{100,j}$ (that has normalized density $\mathrm{CARD}_{\mathrm{temp}, ij}(\vec{x}_d | M_{sat,i}, t_{100,j})$). In this context, the template coefficients $A_{ij}$ represent the fraction of mass contributed to the halo from accretion events in a bin $M_{sat,i},~ t_{100,j}$. We also include an additional template that has uniform density across the full abundance plane; this serves as our ``noise" template. 

We can solve for our linear coefficients by specifying a loss function and optimizing it under a simple set of constraints. If we define a vector of templates as $\overrightarrow{\rm CARD}_{\mathrm{temp}}(\vec{x}_d)$ and a coefficient vector $\vec{A}$, we can rewrite the above expression:

\begin{equation}
   \mathrm{CARD}_{\rm halo, model}(\vec{x}_d) = \overrightarrow{\rm CARD}_{\mathrm{temp}}(\vec{x}_d) \cdot \vec{A}.
\end{equation}
We define our loss function to be the absolute value of the difference between our model CARD density and the true density from the simulations: 

\begin{equation}
    \mathrm{Loss~Function}=|\overrightarrow{\rm CARD}_{\mathrm{temp}} \cdot \vec{A}- \mathrm{CARD}_{\rm halo, true}|.
\end{equation}

We constrain the problem solely by requiring that by requiring that each coefficient $A_{ij} \in [0,1]$ and that the coefficients sum to 1 ($ \sum \vec{A}=\sum_{i} \sum_{j} A_{ij}=1$). We solve for the template coefficients using the \texttt{scipy.optimize.minimize} routine. We emphasize that this model as written contains no knowledge of physics, and could be expanded in a Bayesian framework to incorporate additional information (e.g., the mass of the halo). However, as we show in the next section, even this simple model can be used to recover information about the simulation accretion histories, in some cases with surprising accuracy.

While the modeling method is simple, it relies on a set of template abundance distributions: how best to define and create these templates is less simple. In the next subsection, we introduce our method for creating templates.

\begin{figure*}
    \centering\includegraphics[width=0.95\textwidth]{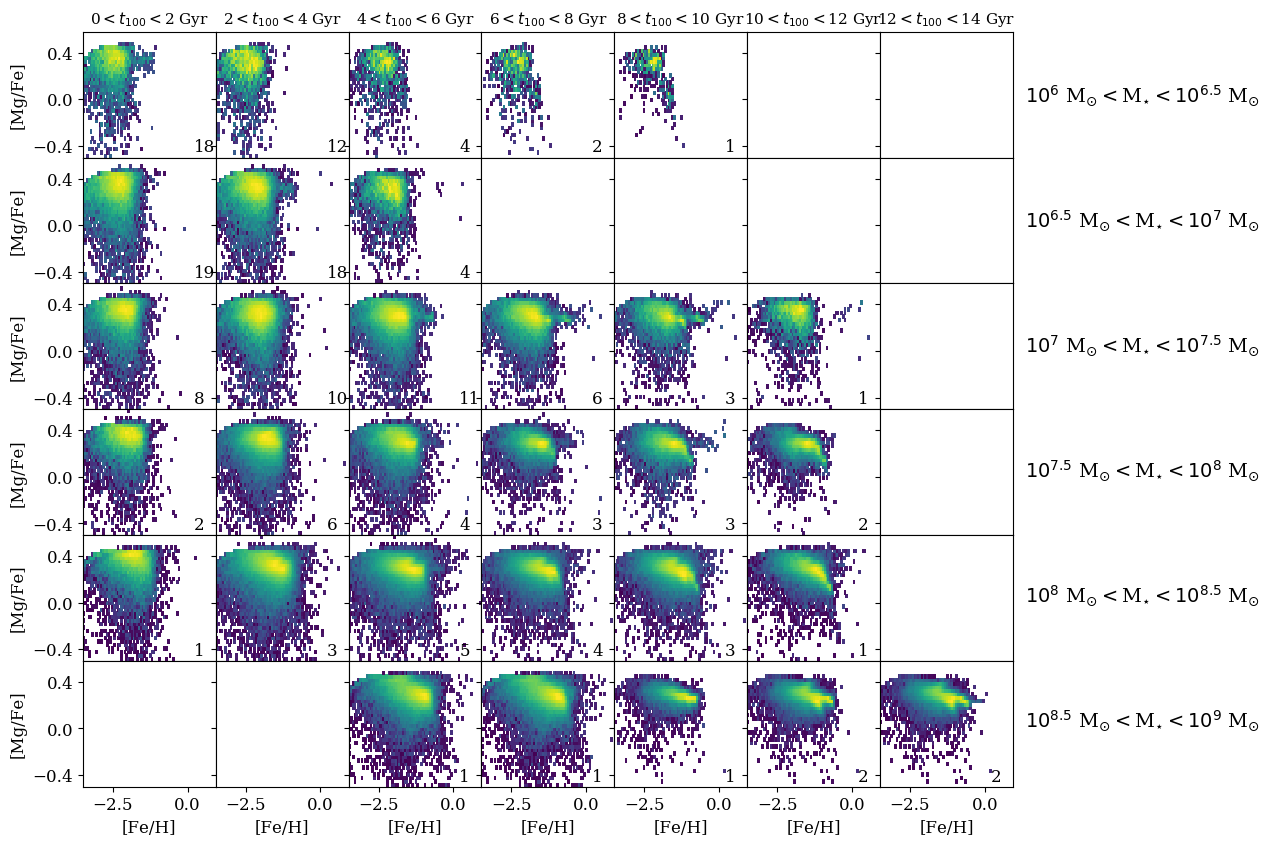}
    \caption{Templates for accretion events (constructed from the streams and phase mixed debris, used to define the stellar halos in this work) for different stellar masses and age ranges. Stellar mass of the event increases from the top row ($10^{6} M_{\odot} < M_{\star}<10^{6.5} M_{\odot}$) to the bottom row ($10^{8.5} M_{\odot}< M_{\star} < 10^{9} M_{\odot}$). Age range, or duration of star formation, increases from left to right. The number in the lower right of each panel indicates the number of dwarf galaxies that were used to construct a given template. We see the general expected trends in the CARDs in these templates; more massive dwarf galaxies have CARDs extending to higher metallicities. At fixed stellar mass, galaxies that assemble more quickly (i.e. have shorter age ranges) have more density in their CARDs at higher [Mg/Fe] values than their counterparts with more extended star formation histories.}
    \label{fig:stpm_tmps}
\end{figure*}

\begin{figure*}
    \centering\includegraphics[width=0.95\textwidth]{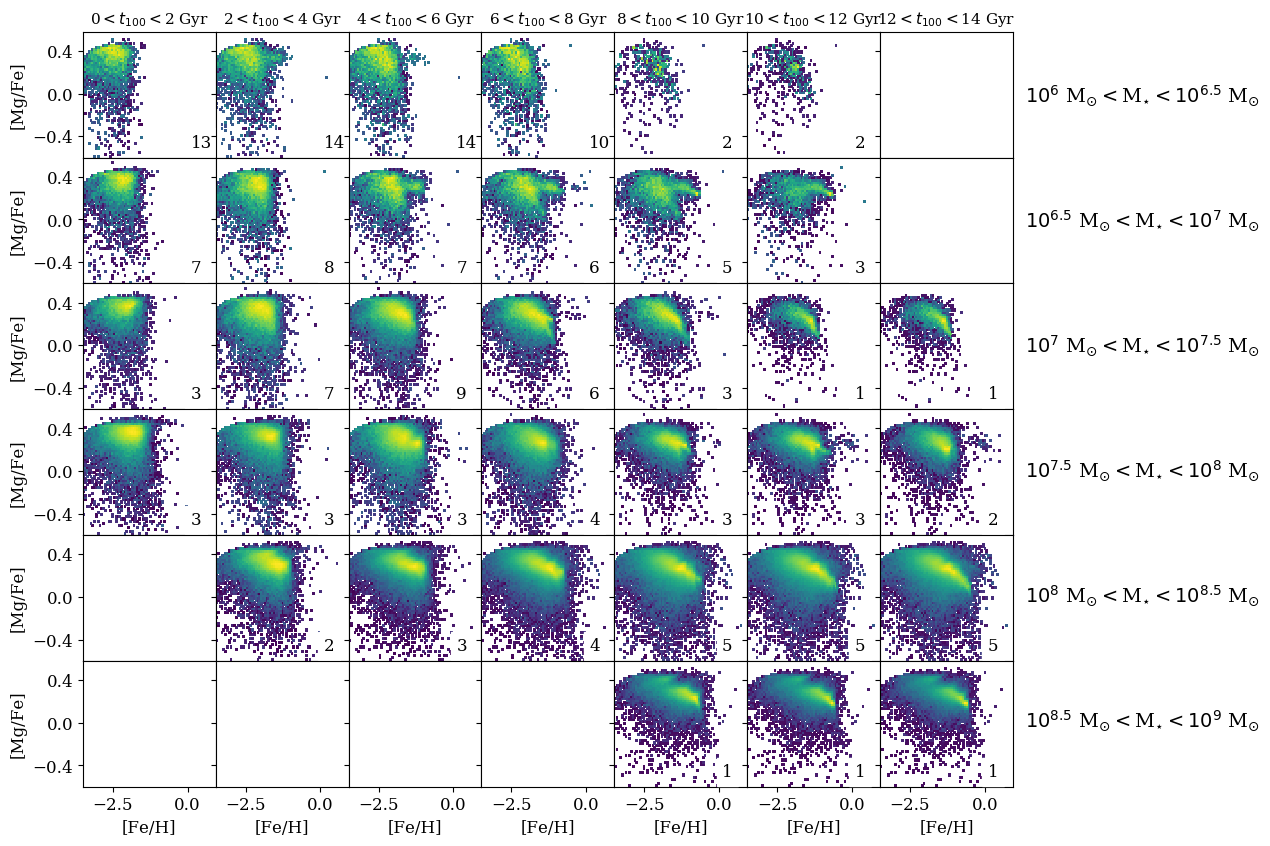}
    \caption{Same as Figure \ref{fig:stpm_tmps}, but for templates constructed from dwarf galaxies.}
    \label{fig:dg_tmps}
\end{figure*}

\begin{deluxetable*}{cl}
\tablecaption{Useful Definitions}
\tablenum{2}
\label{tab:defn}
\tablehead{& Abbreviations and definitions referenced in the text} 

\startdata
CARD &  Chemical abundance ratio distribution; the density of stars in the chemical plane, for two or more elements \\&
(e.g., [Fe/H] vs [Mg/Fe]).\\ 
\hline
Timescales \\
\hline
$t_{\rm form}$ & Formation time of a star particle; time a star particle formed in the simulation, with $t_{\rm form}=0$ defined \\ &  as the beginning of the simulation \\
$t_{100}$ & Quenching time of a halo progenitor or dwarf galaxy. This is equivalent to the formation time of the \\ & youngest star particle associated with a given halo progenitor or dwarf galaxy. \\
\hline
\textit{Templates} \\
\hline
Grid Dimensions & $10^{6} M_{\odot}$ to $10^{9} M_{\odot}$ (in intervals of $10^{0.5} M_{\odot}$) in stellar mass \\ & 0 to 14 Gyr in $t_{100, \mathrm{temp}}$ (in intervals of 2 Gyr)\\
$i$ &  Index for the stellar mass grid dimension \\
$j$ & Index for the quenching time ($t_{100}$). \\
Template Sources & Objects with empirical CARDs used for constructing templates (e.g., present-day dwarfs, or streams phase-mixed debris). \\
CARD$_{\mathrm{temp}, ij}$ & Template abundance distribution for a halo progenitor with $(M_{\star,i}, t_{100,j})$. \\
$t_{100, \mathrm{source}}$ & The quenching time for a halo progenitor or dwarf galaxy used to construct a set of templates. \\
$t_{100, \mathrm{temp},j}$ & The mean quenching time corresponding to a specific CARD$_{\mathrm{temp}, ij}$. \\
$t_{100, \mathrm{temp, max},j}$ ($t_{100, \mathrm{temp, min},j}$)  & The maximum (minimum) quenching time corresponding to a specific CARD$_{\mathrm{temp}, ij}$. \\
\hline
Template Sets \\
\hline
S/PM Templates & Templates constructed using streams and phase-mixed debris as sources. \\ & Default template set is for [Mg/Fe] vs [Fe/H].  Each halo has its own set of S/PM templates, constructed \\
& excluding its own halo progenitors as sources. \\
DG Templates & Templates constructed using present-day dwarf galaxies as sources. \\
\enddata
\end{deluxetable*}

\subsection{Templates}
\label{subsec:temps}

The CARDs modeling method, as described in the previous subsection, relies on a set of template abundance distributions for halo progenitors with different stellar masses and quenching times. In this subsection, we present our method for constructing CARD templates. In our efforts to test the CARDs method in a realistic setting, as well as make the CARDs method ultimately practical for observations, we require a framework for creating templates that is observationally viable. While there are several valid approaches for constructing template CARDs, including from purely theoretical models, here we explore an empirical approach, relying on the abundance distributions of present-day dwarf galaxies within the simulations. We assess this empirical approach within simulations, motivated by the ultimate goal of using observed dwarf galaxy CARDs to recover the properties of the progenitors of the MW and M31 stellar halos. 

Because we are constructing empirical templates, we rely on a catalog of \textit{sources}: the sources for our templates refer to the abundance distributions used to create the templates. We consider two different populations of template sources: 1) halo progenitors (e.g., streams and phase-mixed debris) that contribute to other halos in the simulation suite, and 2) present-day dwarf galaxies (using all available dwarf galaxies from the P21 catalog). By comparing these two catalogs of template sources, we can test the expectation that disrupted dwarfs may be ``better" template sources than present-day dwarfs, and determine if using observed dwarf galaxy CARDs will be viable template sources for the MW and M31 stellar halo progenitors. 

We begin in Section \ref{subsubsec:temp_axes} by discussing how we define the axes of our grid of templates; i.e., the properties of halo progenitors that we are attempting to constrain with this method. In Section \ref{subsubsec:creating_temps}, we then describe our method for creating templates, and show how we can use phase-mixed debris, streams, or dwarf galaxies as template sources. In Section \ref{subsubsec:temp_sets}, we present our CARD templates sets in [Mg/Fe] vs \feh, constructed from streams and phase-mixed debris as well as an additional template set constructed only from present-day dwarf galaxies. Additional templates used in this work (for the [Mg/C] vs \feh distributions) can be found in Appendix \ref{app:temps}. 

\subsubsection{Defining the Grid of Templates}
\label{subsubsec:temp_axes}

Given that we want to use these templates to extract information about the evolution of the halo over time, we first address the question of which timescale we should be using in our template grid. In L15, the two axes of their template grids were stellar mass and accretion time. However, as discussed in P21, assigning an individual accretion time to these events in the context of a cosmological simulation is not straightforward. For example, a dwarf may cross the virial radius of the host galaxy multiple times before becoming tidally disrupted. Furthermore, as can be seen in Figure \ref{fig:form_age_acc_parts}, many of the halo progenitors and dwarf galaxies continue to form stars after falling into the host and completing one or more pericentric passages (a more detailed discussion of satellite quenching will be explored in Samuel et al. 2021, in prep). As a result, in these simulations, the accretion time is not equivalent to the quenching time (as it is in the \citealt{Bullock2005} simulations). Because the quenching time is the more relevant timescale for star formation and chemical enrichment, in this work, we use the quenching time $t_{100}$ as the time dimension for our templates, which we define to be the time a dwarf takes to form 100\% of its stars (defined such that $t=0$ at the beginning of the simulation). This timescale is also equivalent to the formation time of the youngest star particle associated with a given progenitor or dwarf galaxy, and is approximately equivalent to the age range of associated star particles (all sources in the P21 catalog begin forming stars very near the beginning of the simulation). 

In Figure \ref{fig:temp_demo}, we introduce our format for our grid of halo progenitor CARD templates. Each row in Figure \ref{fig:temp_demo} represents a fixed stellar mass range, with stellar mass increasing from the top row (\mstar $\sim 10^6  M_{\odot}$) to the bottom row (\mstar $\sim 10^9 M_{\odot}$). Each column represents a range for $t_{100}$: the far lefthand panels show templates for halo progenitors with $0<t_{100}<2$ Gyr (i.e., templates for halo progenitors that quench within the first two Gyr of the simulation). The far right panels show templates for halo progenitors with $12<t_{100}<14$ Gyr (this can theoretically include halo progenitors that are still forming stars at the present day). The particular templates shown in Figure \ref{fig:temp_demo} are used to demonstrate our procedure for creating templates, which we discuss in detail in the subsequent subsection.

Our grid of templates extends from $10^{6} M_{\odot}$ to $10^{9} M_{\odot}$ (in intervals of $10^{0.5} M_{\odot}$) in stellar mass, and 0 to 14 Gyr in quenching time, $t_{100}$ (in intervals of 2 Gyr). We denote $i$ as the index for stellar mass and $j$ as the index over quenching time. Each grid cell $(i,j)$ shows the template CARD$_{\mathrm{temp}, ij}$ for a halo progenitor with stellar mass $M_{\star, i}$ and quenching time $t_{100, j}$. We note that each grid cell $(i,j)$ is actually representing a range of stellar mass and $t_{100}$, and as such has a corresponding $M_{\star, \mathrm{min}, i}, M_{\star, \mathrm{max}, i}$ as well as a $t_{100, \mathrm{min}, j}, t_{100, \mathrm{max}, j}$. However, for convenience in notation, we generally refer to a given grid cell with indices $(i,j)$ as containing the template CARD$_{\mathrm{temp}, ij}$ for a halo progenitor with properties $(M_{\star, i}, t_{100, j})$.

 \subsubsection{Creating Templates}
 \label{subsubsec:creating_temps}
 
 The observed CARDs from present-day dwarf galaxies in the LG provide one avenue for constructing empirical CARD templates for MW and M31 halo progenitors. One challenge, as discussed in Section \ref{sec:p21_cat} and shown in Figure \ref{fig:mstar_deltat}, is that dwarf galaxies occupy a different region in the $(M_{\star}, t_{100})$ plane from halo progenitors (though there is substantial overlap). On average, at fixed stellar mass, dwarf galaxies will have longer quenching times than the phase-mixed debris and stellar streams, as a result of the fact that the surviving dwarf galaxies are not quenched as a result of being accreted by the host. If we are to construct empirical template sets from the observations, we will want to use the abundance distributions from local low mass dwarfs in creating our templates. Given that dwarf galaxies tend to have larger $t_{100}$ than streams and phase-mixed debris (as seen in Figure \ref{fig:mstar_deltat}), how can we use dwarf galaxy CARD templates to recover the properties (including $t_{100}$) of the halo progenitors? 

In order to compare the abundance distributions of dwarf galaxies and halo progenitors, we have to account for the fact that, at fixed stellar mass, the $t_{100}$ distributions for these two populations are not the same. Therefore, we define two versions of $t_{100}$:

\begin{itemize}
\item $t_{100, \mathrm{source}}$: the time a template source (either a halo progenitor or present-day dwarf galaxy) from the P21 catalog takes to form 100\% of its stars. This is the quantity plotted on the $y$-axis in Figure \ref{fig:mstar_deltat}.
   \item $t_{100, \mathrm{temp}}$: the time a halo progenitor represented by specific CARD$_{\mathrm{temp}, ij}$ template takes to form 100\% of its stars. This is the quantity shown in Figure \ref{fig:temp_demo} (as well as subsequent Figures containing CARD templates).
\end{itemize}
 
 Figure \ref{fig:temp_demo} demonstrates the structure of our template grid, with stellar mass increasing from the top row to the bottom row and $t_{100, \mathrm{temp}}$ increasing from left to right. To construct our templates, we begin by specifying a column $j$, and work from left to right. Given a specified column $j$, we then work with one source at a time. For each object in the source catalog: 

\begin{enumerate}
\item We select all star particles that formed before $t_{100,\mathrm{temp, max},j}$. For example, to make the leftmost column of templates, which are for accretion events with $0<t_{100, \mathrm{temp}}<2$ Gyr, we use all the star particles from a given accretion event that formed in the first 2 Gyr of the simulation.   

\item We compute the stellar mass of the subset of star particles with $t_{\rm form}<t_{100,\mathrm{temp, max},j}$. Based on this stellar mass, the accretion event is assigned to a stellar mass bin $i$ (or row in Figure \ref{fig:temp_demo}).

\item We compute the CARD for the subset of star particles with $t_{\mathrm{form}}<t_{100,\mathrm{temp, max},j}$. This CARD is then averaged with the CARDs from other template sources that contribute to the same bin $(i,j)$, to create the CARD$_{\mathrm{temp},ij}$.

\end{enumerate}

Once we have assigned each source to a bin $(i,j)$, we compute the average normalized density of all accretion events that contribute to each bin to create the template CARD$_{\mathrm{temp},ij}$. The distributions from individual events are normalized before averaging so that the most massive events do not dominate the average densities. We then repeat this procedure for each column (i.e., the next interval for $t_{100, \mathrm{temp}}$). We note that once the template source is no longer forming stars (due to tidal disruption, quenching, etc.), it no longer contributes to templates. Accretion events only contribute to templates with $t_{100, \mathrm{temp}} \leq t_{100, \mathrm{source}}$.

Figure \ref{fig:temp_demo} demonstrates our procedure for creating templates, using two sources from the P21 catalog: a present-day coherent stream ($M_{\star}=10^{7.15} M_{\odot}, t_{100, \mathrm{source}}=10.0$ Gyr; outlined in pink) and a present day dwarf galaxy ($M_{\star}=10^{8.7} M_{\odot}, t_{100, \mathrm{source}}=13.25$ Gyr; outlined in blue). While we only used two sources, as a result of our procedure for creating templates, each source contributes to multiple templates on the grid in Figure \ref{fig:temp_demo}. The present day dwarf galaxy contributes to seven different templates: one for each time interval (as the dwarf galaxy is forming stars over the full duration of the simulation) and all with stellar mass $M_{\star}>10^{7.5}$ \msun. As the dwarf's stellar mass increases over time, it contributes to templates in higher stellar mass bins. In contrast, the present day stream contributes to only five templates, all with $M_{\star}<10^{7.5}$ \msun: because the stream has $t_{100}=10.0$ Gyr, it no longer contributes to templates beyond the $8<t_{100, \mathrm{temp}}<10$ Gyr time interval (i.e., it doesn't contribute to the two rightmost columns in Figure \ref{fig:temp_demo}).

The template set shown in Figure \ref{fig:temp_demo} is purely for illustrative purposes: a template set created from only two sources is obviously of limited usefulness. In the subsequent subsubsection, we introduce the various template sets used for analysis in this paper. 
 
\subsubsection{Template Sets}
\label{subsubsec:temp_sets}

Using the procedure outlined in the previous section, we use the P21 catalog to create four primary sets of templates: two template sets for the distributions in [Mg/Fe] vs \feh~and two template sets for the distributions in [Mg/C] vs [Fe/H]. We choose to include carbon in addition to magnesium because carbon predominantly traces the enrichment due to stellar winds, therefore incorporating information about an additional source of chemical enrichment (on a different timescale). For each combination of elements, we create one template set from the material that has been phase-mixed or has formed a stream in the isolated MW simulations (i.e., the accretion events whose properties we are trying to recover) as well as dwarf galaxies at present day.

Figure \ref{fig:stpm_tmps} shows the resulting CARD templates in [Mg/Fe] versus [Fe/H] created from all of the phase-mixed debris and coherent streams from the P21 catalog in the \textit{Latte} suite. These templates are hereafter referred to as the S/PM templates. Figure \ref{fig:dg_tmps} shows the [Mg/Fe] vs [Fe/H] templates created from the surviving dwarf galaxies (from both the \textit{Latte} and ELVIS on FIRE suites), hereafter referred to as the DG templates. In both Figures \ref{fig:stpm_tmps} and \ref{fig:dg_tmps}, as in Figure \ref{fig:temp_demo}, templates increase in stellar mass from the top row (\mstar $\sim 10^6 M_{\odot}$) to the bottom row (\mstar $\sim 10^9  M_{\odot}$). In constructing the templates, we average over as many accretion events with the desired properties that we have available; the number of events contributing to each template is indicated by the number in the lower right corner of each panel in Figures \ref{fig:stpm_tmps} and \ref{fig:dg_tmps}. In this work, we also make use of template sets constructed from [Mg/C] vs [Fe/H]; these template sets are shown in Appendix \ref{app:temps}. In Appendix \ref{app:temps}, we also discuss several additional template sets that we tested, including ``master" templates (utilizing halo progenitors and dwarf galaxies as sources) as well as ``early" and ``late" templates (dividing template sources by $t_{100}$). We refer the reader to Appendix \ref{app:temps} for further details. 

As noted in Section \ref{subsec:temps}, we create a distinct S/PM template set for each individual halo, which excludes all of that halo's progenitors (where each simulation's ``stellar halo" is defined as the ensemble of star particles belonging to streams and phase-mixed debris identified in P21). Therefore, we note that the template set shown in Figure \ref{fig:stpm_tmps}, which includes all of the P21 \textit{Latte} halo progenitors, is not used for analysis. We choose to exclude a given halo's progenitors from its template set in order to better directly compare the performance of S/PM templates with the templates created from present-day dwarf galaxies. If a given halo's progenitors are included in the S/PM templates, the S/PM templates will always perform better than the dwarf galaxy templates by design. 

Both template sets show the same general trends as expected from the chemical enrichment models. As stellar mass increases, the CARDs extend to higher and higher iron metallicities. At fixed stellar mass, templates with shorter age ranges (i.e., smaller $t_{100}$, on the lefthand side of the grid of templates) have higher density in their CARDs at higher [Mg/Fe], whereas templates with larger age ranges (i.e., high $t_{100}$, on the righthand side of the grid) have higher density at lower [Mg/Fe]. This results from the longer timescales associated with Type Ia SNe (the primary source of iron enrichment) with respect to Type II SNe (the primary source of $\alpha$-element enrichment).

While the expected general trends are found in both template sets, there are a few salient differences between the two template sets worth noting. First of all, the fact that present-day dwarf galaxies are ``temporally biased" with respect to their disrupted counterparts can be identified based the low number of S/PM templates with long age ranges relative to the DG templates (i.e., the grid in Figure \ref{fig:stpm_tmps} is relatively sparsely populated on the righthand side compared to the grid in Figure \ref{fig:dg_tmps}). Therefore, had we constructed the template sets based on the present-day properties alone, we would have found that many of the S/PM accretion events would not have been found within the parameter space of the present-day DG templates. 

In addition, we note that this technique relies upon the assumption that dwarf galaxies and halo progenitors with similar stellar masses and age ranges will have similar CARDs; we assume that the CARDs within our template set will be representative of the halo progenitor CARDs. Comparing the different CARDs on Figures \ref{fig:stpm_tmps} and \ref{fig:dg_tmps}, we can immediately see that not all of the distributions within the same gridpoints are identical. We explore some of these differences in more detail in Section \ref{sec:discussion}.

We emphasize that these templates are empirical templates derived from the FIRE simulations and are designed to be used on the FIRE simulations. \textbf{The templates from the FIRE simulations should not be applied to observational data.} While the overall properties of the simulations make them excellent testbeds for studying galaxy formation phenomena, it remains important to keep in mind that the abundances are not literal abundances, but rather labels in a simulation. As discussed in Section \ref{sec:fire_sims}, the average \feh of the stars in the satellites in these these simulations are too low compared to the observed dwarf galaxies (\citealt{Escala2018}, \citealt{Wheeler2019}, P21), likely as a result of the assumed delay-time distributions for SNe Ia. In addition, the harsh ridge seen in all the CARDs at high Mg/Fe results because of the initialization: star particles are initialized with \feh $=-4$ and \afe $=0$, resulting in a hard upper limit for the $\alpha$ abundances at a given iron abundance. Both of these particular issues will be addressed with the new FIRE-3 physics model (Hopkins et al., in prep). 
Therefore, while these templates can be used in estimating the assembly histories of the \textit{Latte} galaxies, we emphasize that an empirical template set for use on observational data should be constructed from observational data (or from theoretical models calibrated to reproduce the observed abundance distributions).

\section{Results}
\label{sec:results}

In this section, we discuss the results of using CARDs to estimate the assembly histories of the stellar halos from the \textit{Latte} suite of simulations. We discuss the results from two simulations in detail: one in which the technique performs very well (m12f) and one for which the technique does not accurately recover the assembly history (m12c). 

For our two example cases, in Section \ref{subsec:res_cards}, we first explore how well our four template sets are able to reproduce the stellar halo CARDs. In Section \ref{subsec:res_massspec}, we discuss how well the CARDs technique recovers the mass spectrum of accreted dwarfs. In Section \ref{subsec:res_assembly}, we demonstrate how well we can recover the mass accreted as a function of progenitor quenching time. In Section \ref{sec:res_sum}, we summarize the results for our analysis for all the simulations. For figures and detailed discussion of the remaining five simulations not discussed in this section, we refer the reader to Appendix \ref{app:detailed_res}.

\subsection{Model CARDS}
\label{subsec:res_cards}

In assessing the performance of this technique, we begin by examining the results from what we are modeling directly: the simulation abundance distributions.

In this section, we compare the CARDs derived from our modeling procedure with the true simulation CARDs, for the two simulations we discuss in detail. We first focus on our success case: m12f. The CARDs from the accreted star particles (as defined in Section \ref{sec:p21_cat}) are shown in the far left hand panels of Figure \ref{fig:m12f_dens}; we model both [Mg/Fe] vs \feh~(top panels) as well as [Mg/C] vs \feh~(lower panels). The resulting model CARDs, using our stream/phase mixed templates (second from the left) and dwarf galaxy templates (middle panels), appear to be a good representation of the data. This is quantified by the model residuals, as shown in the far right panels; the value of the residuals (defined here as model$-$data) are indicated by the colorbar. We see that for this simulated galaxy, residuals are quite low, and the CARDs constructed from all four sets of templates match the halo CARDs very well.

In contrast, Figure \ref{fig:m12c_dens} shows the CARDs results for m12c. Even by eye, the fact that the model CARDs are not good representations for the simulation CARDs is immediately apparent. The CARD for m12c shows high density in near \feh $\sim -1$ and [Mg/Fe]$\sim 0.2$. This feature, which arises from the most massive accretion event in m12c ($M_{\star}=10^{8.7}$ and $t_{100}=12.35$ Gyr), is not captured by the templates; as a result, the model strongly underpredicts the density in this region of the CARD. Unsurprisingly, given the poor fit to the data, the CARDs technique does not recover the assembly history correctly. 

The inability of the method to accurately reproduce the CARD of m12c highlights the limitations of a fundamental assumption of the model: that accretion events with similar $M_{\star}, t_{100}$ will have similar CARDs. We discuss the diversity in the formation histories (and therefore the CARDs) of the most massive dwarfs in our sample in detail in Section \ref{sec:discussion}. 

\subsection{Mass Spectra of Accreted Dwarfs}
\label{subsec:res_massspec}

Having assessed the fit of our model CARDs, we turn now to the ability of the technique to recover the assembly histories. We emphasize again that the model presented in this paper contains no information about mass or time: the model only uses the simulation CARDs and the CARDs from the templates. However, using the stellar mass and $t_{100}$ labels for each of our templates, we can assess how well the technique recovers the assembly histories of the simulated galaxies (which are modeled indirectly). We consider the mass spectrum of accreted dwarfs (i.e., the fraction of mass contributed to the halo as a function of the mass of the progenitor) and the stellar mass accreted as a function of time (where we again use $t_{100}$ instead of ``accretion time").

Figure \ref{fig:m12fc_mass} shows the mass spectra for the simulations m12f (left panels) and m12c (right panels), with the true mass spectra shown as gray squares connected by the gray lines. The derived mass spectra from the models using the S/PM templates, the DG templates, the [Mg/C] S/PM templates and the [Mg/C] DG templates are shown in pink, blue, peach, and green, respectively. The top panels shows the mass spectrum plotted linearly in mass fraction; the middle left panel shows the results logarithmically in mass fraction. The lowest panel shows the residuals from each model, as well as the RMS dispersion of the residuals as the dashed lines.

It is immediately apparent that the mass spectrum of accreted dwarfs is recovered extremely accurately for m12f, using all template sets. The RMS dispersion of the residuals is 0.05 for the stream/phase-mixed templates, 0.04 for the dwarf galaxy templates in [Mg/Fe] vs \feh; 0.09 for the [Mg/C] vs [Fe/H] S/PM templates; and 0.03 for the templates for [Mg/C] vs \feh. While the logarithmic panel highlights that the dwarf galaxy templates overestimate the contribution of the lowest mass bin (finding $\sim$ 3--5 \% contribution), overall, the CARDs modeling technique recovers the mass spectrum very accurately, with the [Mg/C] DG templates recovering the true mass spectrum to the highest precision. We also note that a very low mass contribution from high mass accretion events is predicted for both sets of S/PM templates; given that this is unphysical (as a low contribution from a $10^9 M_{\odot}$ accretion event implies a very large halo mass), this could be addressed by constructing a more complex model incorporating prior information on the total mass of the halo and modeling the number of accretion events instead of the mass fractions.  

In contrast, the righthand panels of Figure \ref{fig:m12fc_mass} show the mass spectrum results for the m12c analysis. Unsurprisingly (based on the poor fit to the simulation CARD), the true mass spectrum is not well recovered. All template sets perform equally poorly at recovering the mass spectrum. Because the dwarf progenitor of the most massive accretion event in m12c has high density at \feh $\sim -1$, and does not continue to enrich to higher \feh and lower \afe as would be expected, the modeling procedure derives a mass spectrum where the dominant event occurs in a lower mass bin. 

We see from the m12f results that this technique is capable of accurately recovering the mass spectra of accreted dwarf galaxies. For 4/7 of the \textit{Latte} suite halos (m12b, m12f, m12m, and m12w), we are able recover the mass spectrum to a precision of 10\% or less, using the DG and/or the [Mg/C] DG templates. However, for the remaining 3/7 galaxies (m12c, m12i, and m12r), the mass spectra are not as accurately recovered: this occurs, as in m12c, when the CARDs from the halo accretion events are not well represented by the template set.

\subsection{Mass Assembly versus Quenching Time}
\label{subsec:res_assembly}

In the previous subsection, we discussed how well the CARDs method recovered the mass accreted as a function of the mass of the progenitor (i.e., the mass spectrum of accreted dwarfs). In this section, we look at the fraction of mass accreted as a function of the progenitor's quenching time. Figure \ref{fig:m12fc_time} shows the accreted stellar mass fraction as a function of progenitor $t_{100}$ for m12f (left) and m12c (right). Because we are using $t_{100}$ as a proxy for what is traditionally thought of as accretion time, we show the cumulative distributions, though the residuals in the lower panels of Figure \ref{fig:m12fc_time} are shown for the mass fractions and not the cumulative mass fractions, as in Figure \ref{fig:m12fc_mass}. 

Based on the lower panels of Figure \ref{fig:m12fc_time}, it is immediately apparent that the residuals for the mass fraction as a function of $t_{100}$ are significantly larger than for the mass spectra. Even for m12f, the residuals have an RMS dispersion of $\sim 0.2-0.3$; while the distribution of mass accreted as a function of $t_{100}$ is not recovered as accurately as the mass spectrum, the model still does recover the fact that most of the mass is from accretion events with intermediate $t_{100}$ (i.e., $3~\mathrm{Gyr}~<t_{100}<7$ Gyr). 

However, as with the mass spectrum, the inferred mass assembly over time for m12c is in poor agreement with its actual assembly history. Because of the anomalous CARD for the most massive accretion event in this halo, which doesn't enrich beyond \feh $\sim -1$ and [Mg/Fe]$\sim 0.2$, the model created from the templates favors an earlier assembly history than the galaxy actually experienced.

While generally the mass spectrum is recovered with higher accuracy than the mass assembled as a function of time, there are a number of prospects for potential improvements to the methods that could be explored in future work. For example, in this work, we have focused exclusively on $t_{100}$ in creating templates and quantifying the properties of halo progenitors; however, it's possible that other timescales, such as $t_{90}$ or the interquartile age range, could improve the inferences. Exploring the moments of the abundance distributions, as well as incorporating additional elements tracking different sources of chemical enrichment, could also prove useful in constraining the properties of the progenitors.

However, as with m12c in this work, if the template CARDs are not a good representation of the halo progenitor CARDs, the inferences for the assembly histories derived using this method will not be correct. Fortunately, our first clue that there is a mismatch between the halo progenitor CARDs and the template CARDs are the high residuals in comparing the model CARDs with the data. In the next section, we summarize the results of modeling all seven of the halos from the \textit{Latte} suite used in this work, and demonstrate the connection between the residuals in the CARDs models and the residuals in the inferred assembly histories.

\subsection{Summary of All Simulation Results}
\label{sec:res_sum}

While we show the detailed results for the assembly histories of the individual simulations in Appendix \ref{app:detailed_res}, Figure \ref{fig:results} summarizes the residuals from modeling the CARDs of all of the \textit{Latte} halos. The top panels demonstrate the results from inferring the mass spectrum of accreted dwarfs, and the lower panels show the results from inferring the fraction of mass accreted as a function of progenitor $t_{100}$. In all panels, each color represents the results from a different simulation. Circular points show the results from using the [Mg/Fe] vs [Fe/H] template sets, while square points indicate results from using the [Mg/C] vs [Fe/H] template sets. Open symbols show the results from using the S/PM templates, and filled points show the results from the DG templates. 

\begin{figure*}
    \centering
    \includegraphics[width=\textwidth]{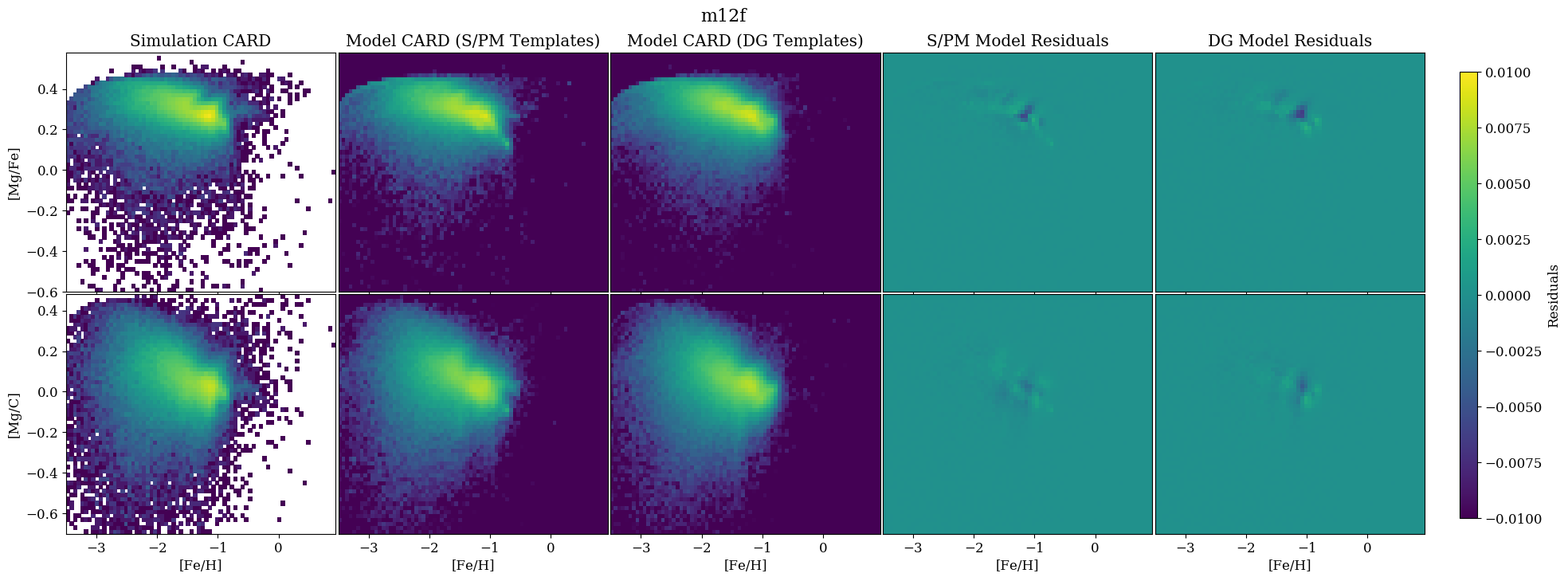}
    \caption{CARDs results for the simulation m12f. The true simulation CARDs (using the streams and phase-mixed debris identified in P21) are shown in the far left panels, with [Mg/Fe] vs \feh~ plotted in the top row and [Mg/C] vs \feh shown in the bottom row. The resulting model CARDs using the S/PM templates are shown in the second from the left panels, and the CARDs using the DG templates are in the center. Residuals from all models are shown in the righthand panels; the magnitude of the residuals is shown by the colorbar. For the case of m12f, the models utilizing all of the different template sets produce a good fit to the data, as demonstrated by the very low residuals.}
    \label{fig:m12f_dens}
\end{figure*}

\begin{figure*}
    \centering
    \includegraphics[width=\textwidth]{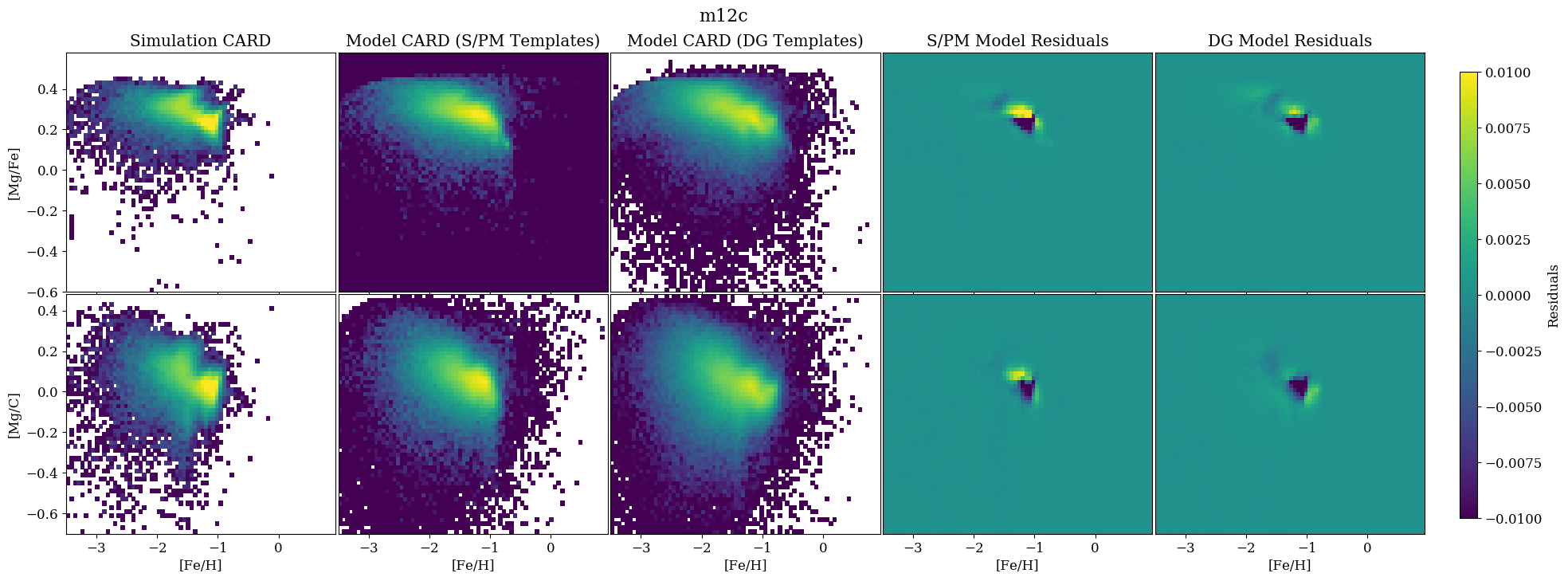}
    \caption{Same as Figure \ref{fig:m12f_dens}, but for the simulation m12c. In contrast to the m12f results, none of the models shown here are a good fit to the data: the high density feature in the simulation CARD at \feh$\sim-1$ is not well represented by the templates. As a result, all templates result in model CARDs with very high residuals.}
    \label{fig:m12c_dens}
\end{figure*}

\begin{figure}
 \centering
    \includegraphics[width=0.5\textwidth]{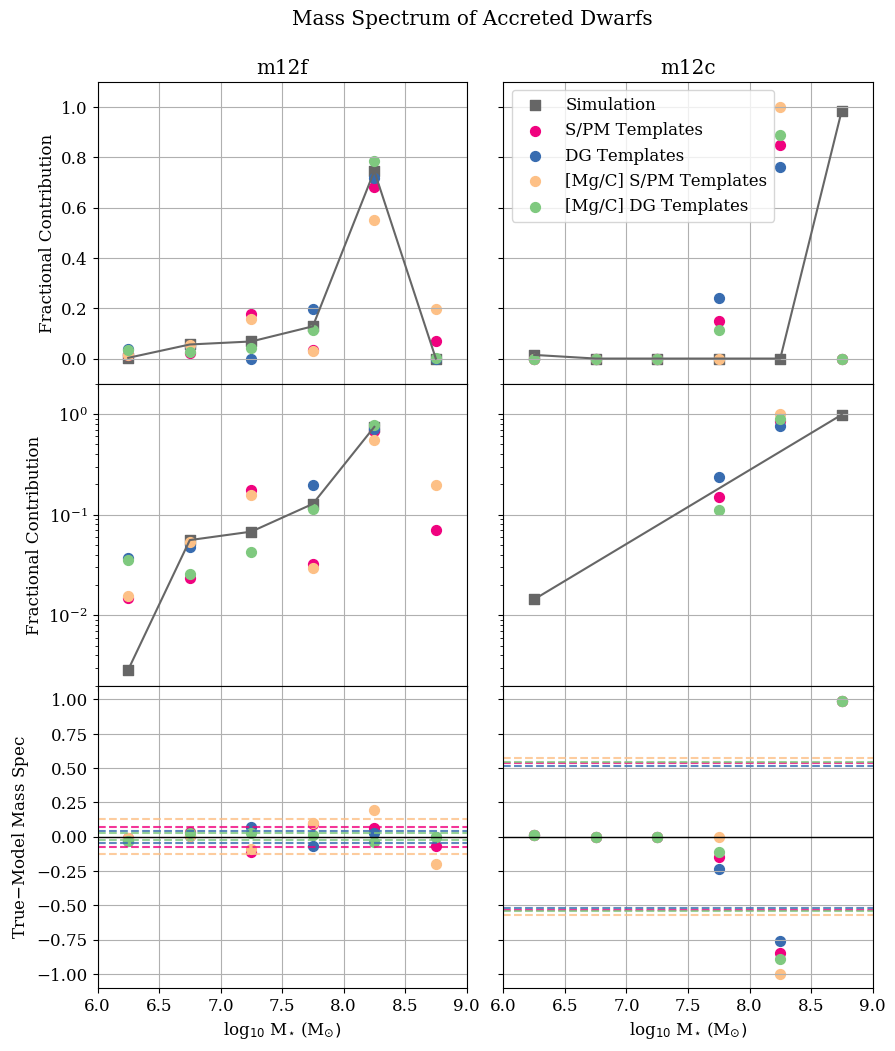}
    \caption{Mass spectra of accreted dwarfs as derived from the different template sets, for the best case (m12f) and a failure case (m12c). Top panels show the mass fraction as a function of progenitor stellar mass on a linear scale; middle panels show the mass fractions on a logarithmic scale. The true mass spectrum from the simulations are shown by the grey squares. The resulting mass spectra derived using the stream/phase mixed templates, dwarf galaxy templates, the S/PM templates for the [Mg/C], and the dwarf galaxy templates for the [Mg/C] distributions are shown by the pink, blue, peach and green points, respectively. Lower panels show the residuals from the results derived from each set of templates, with the RMS dispersion for each fit indicated by the dashed lines. The results for m12f are shown on the left and the results for m12c are shown on the right. } 
    \label{fig:m12fc_mass}
\end{figure}

\begin{figure}
    \centering
    \includegraphics[width=0.5\textwidth]{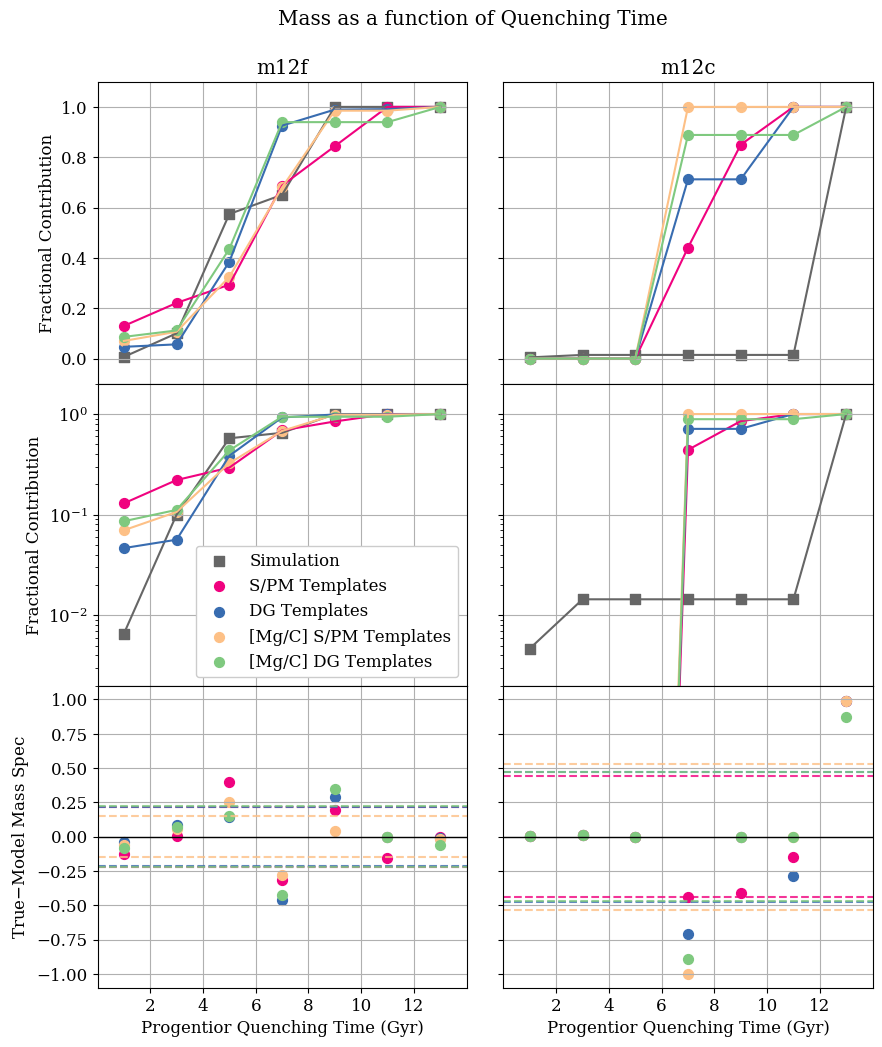}
    \caption{Cumulative mass fraction as a function of $t_{100}$, for the success case (m12f, left) and failure case (m12c, right).}
    \label{fig:m12fc_time}
\end{figure}

The $x-$axes on the left panels of Figure \ref{fig:results} indicates the RMS dispersion of the residuals from the CARDs models (i.e., the RMS dispersion of the residuals plotted in the righthand panels of Figures \ref{fig:m12f_dens} and \ref{fig:m12c_dens}). 
The $y-$axes of the top panels show the RMS dispersion of the residuals for the mass spectra, and the $y-$axes of the lower panels show the RMS dispersion for the mass assembly histories.

This figure demonstrates the following key results:

\begin{enumerate}
    \item \textbf{The CARDs method infers the mass spectrum of accreted dwarfs to a precision of $10\%$ or better using our DG templates for four out of the seven simulations (m12b, m12f, m12m, and m12w).} The top left panel of Figure \ref{fig:results} shows clustering of points in the lower lefthand corner: for these simulations, the CARD has been modeled accurately (i.e., with low residuals), and the mass spectrum has also been recovered accurately. While the m12w CARD model has high residuals, its mass spectrum is still inferred relatively accurately for the DG templates and the [Mg/C] templates; see Appendix \ref{app:detailed_res} for detailed discussion of the individual simulation. 
    \item \textbf{The fraction of mass as a function of $t_{100}$ is generally inferred less accurately than the mass spectrum, with a precision on the order of $\sim 20-30 \%$ for five out of the seven simulations (m12b, m12f, m12i, m12m, and m12r).} Comparing the top left panel with the bottom left panel of \ref{fig:results} shows that the residuals are higher for the mass accreted as a function of quenching time than for the mass spectrum. The only exception to this is m12r; see Appendix \ref{app:detailed_res} for further discussion. 
    \item \textbf{Our first indication that the assembly history will not be inferred accurately is the quality of the fit to the halo CARD.} The $x-$axis on the left panels of Figure \ref{fig:results} indicates the RMS dispersion of the residuals from the CARDs models (i.e., the RMS dispersion of the residuals plotted in the righthand panels of Figures \ref{fig:m12f_dens} and \ref{fig:m12c_dens}). We see a correlation between the RMS dispersion in the CARDs residuals and the RMS dispersion of the parameters of the accretion history (the exception being the mass spectrum inference for m12w; see appendix \ref{app:detailed_res} for discussion). As discussed in the previous section, if the templates are not representative of the halo accretion events, the model CARDs will be a poor fit to the true CARDs and the derived accretion histories will not be correct.
    \item \textbf{The DG templates perform comparably to, if not better than, the S/PM templates.} In the middle panel of Figure \ref{fig:results}, we directly compare the results of our modeling procedure using DG templates ($y-$axis) versus S/PM templates ($x-$axis). The fact that most of the points lie below the 1:1 line in the top middle panel of Figure \ref{fig:results} demonstrates that the DG templates infer the mass spectrum better than the S/PM templates for the majority of the simulations, for both [Mg/Fe] vs [Fe/H] template sets as well as [Mg/C] versus [Fe/H]. This is likely a result of the fact that the star particle membership assignments are the cleanest in the P21 dwarf galaxies, as compared with phase-mixed debris. Even in the simulations, identifying stars affiliated with a common progenitor in the phase-mixed halo without contamination from the disk or other progenitors is challenging. Further cleaning of phase-mixed debris samples in the P21 catalog could result in improved performance of the S/PM templates.    
    \item \textbf{The inclusion of information from Carbon (which traces stellar wind enrichment) in the [Mg/C] templates marginally improves the inferences of the mass spectrum, but doesn't appear to improve the inferences of the quenching times.} The right panels of Figure \ref{fig:results} compare the results from using the Mg/C templates versus the Mg/Fe templates. The top panel demonstrates that the inferences of the mass spectrum for both template sets are comparable, with most results lying near the 1:1 line. However, the Mg/C templates do infer the mass spectrum slightly more accurately on average. The results for mass accreted vs $t_{100}$ are comparable for the the two template sets, with the Mg/Fe templates performing slightly better on average. 
\end{enumerate}

What gives rise to our less accurate inferences? If the template CARDs are not representative of the halo progenitors, our inferences about the assembly history will not be accurate. In particular, Figure \ref{fig:results} highlights that the inferences for m12c and m12w are usually the worst: while both of these simulations experience a massive, late time accretion event, their CARDs are very different. In the next section, we look more closely at the four most massive sources in the P21 catalog in detail. For the results of modeling the CARDs excluding accretion events with $M_{\star}>10^8 M_{\odot}$, we refer the reader to Appendix \ref{app:lower_mass}.

\begin{figure*}
    \centering
    \includegraphics[width=0.9\textwidth]{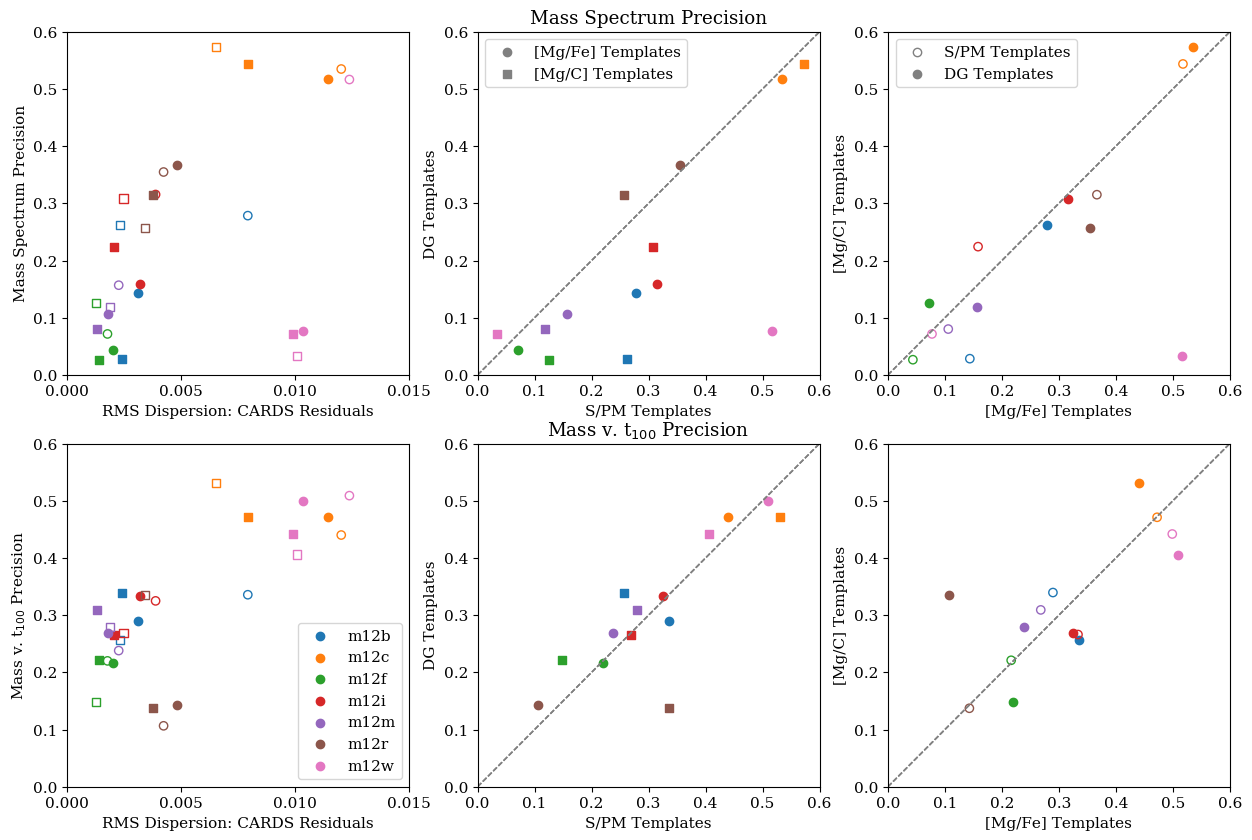}
    \caption{Summary of our results for all simulations. Colors across all panels correspond to results from specific simulations. Top panels show the resulting precision from inferring the mass spectrum of accreted dwarfs, and bottom panels show resulting precision from inferring mass as a function of $t_{100}$. \textit{Left panels:} inferred precision of mass spectrum and $t_{100}$ versus the RMS dispersion of the model CARDS. \textit{Middle panels:} comparison of the results from the DG templates vs the S/PM templates. Results from the [Mg/C] templates are shown as squares while the results from the [Mg/Fe] templates are shown as circles. The DG templates perform comparably or better than the S/PM templates. \textit{Right panels:} Comparison of the results from the [Mg/C] templates versus the [Mg/Fe] templates. S/PM template results are shown as open symbols, while DG template results are filled circles. We see some marginal improvement in the inferred mass spectra by including [Mg/C], though no real improvement in the inferred quenching times.}
    \label{fig:results}
\end{figure*}

\section{Discussion}
\label{sec:discussion}

\subsection{Evolutionary Histories of Massive Dwarfs}
In the previous section, we discussed at length that the simulation m12c's CARD and assembly history were poorly recovered, largely due to the fact that m12c's most massive accretion event ($M_{\star}=10^{8.7} M_{\odot}$, $t_{100}=12.35$ Gyr) has a CARD that is not well represented by the templates. A fundamental assumption of the CARDs modeling method presented here and in L15 is that halo progenitors with the same $(M_{\star}, t_{100})$ should have similar CARDs. Why does the most massive accretion event in m12c have such an odd abundance distribution for a stream with its mass and $t_{100}$? Here, we discuss in more detail the chemical enrichment histories of the most massive streams and dwarf galaxies in our sample, to better understand the diversity of observed CARDs.

\begin{figure}
    \centering
    \includegraphics[width=0.5\textwidth]{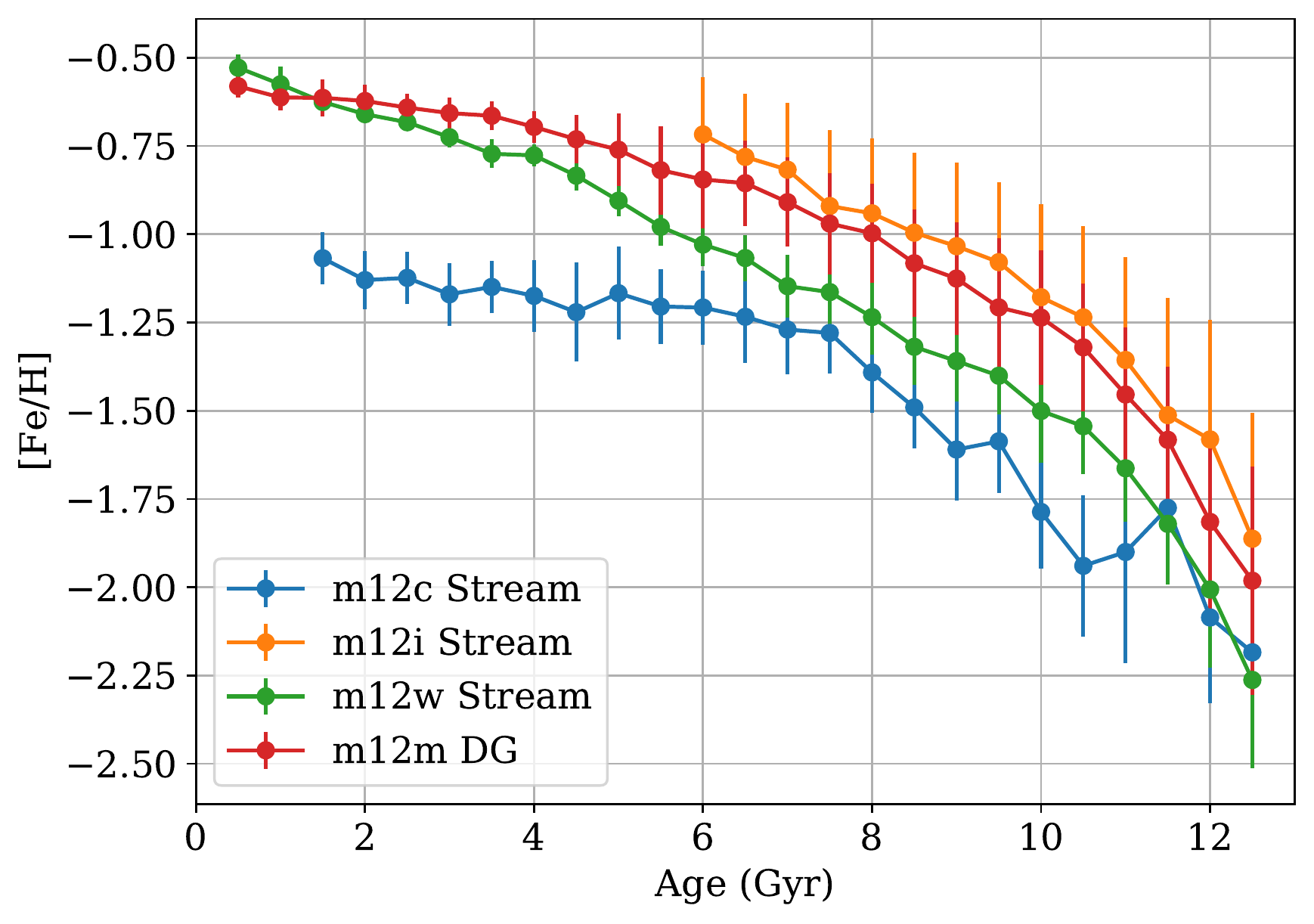}
    \caption{Mean iron abundance as a function of stellar age for the massive streams and dwarf galaxy in the P21 catalog. Errorbars indicate the sample variance in each age bin. The iron abundance of m12c is lower than the other dwarfs/halo progenitors in its mass range at all stellar ages. }
    \label{fig:age_v_fe}
\end{figure}

The P21 catalog contains four objects that have stellar masses $M_{\star}>10^{8.5} M_{\odot}$; massive streams in m12c, m12w, and m12i, as well as a present-day dwarf galaxy in m12m (the only source for the most massive template for the DG templates). Figure \ref{fig:age_v_fe} shows the mean metallicity of particles for each of these massive objects as a function of age; the stream in m12c (shown in blue) is significantly more metal poor at all times than the other massive streams and dwarf galaxy.

\begin{figure}
    \centering
    \includegraphics[width=0.5\textwidth]{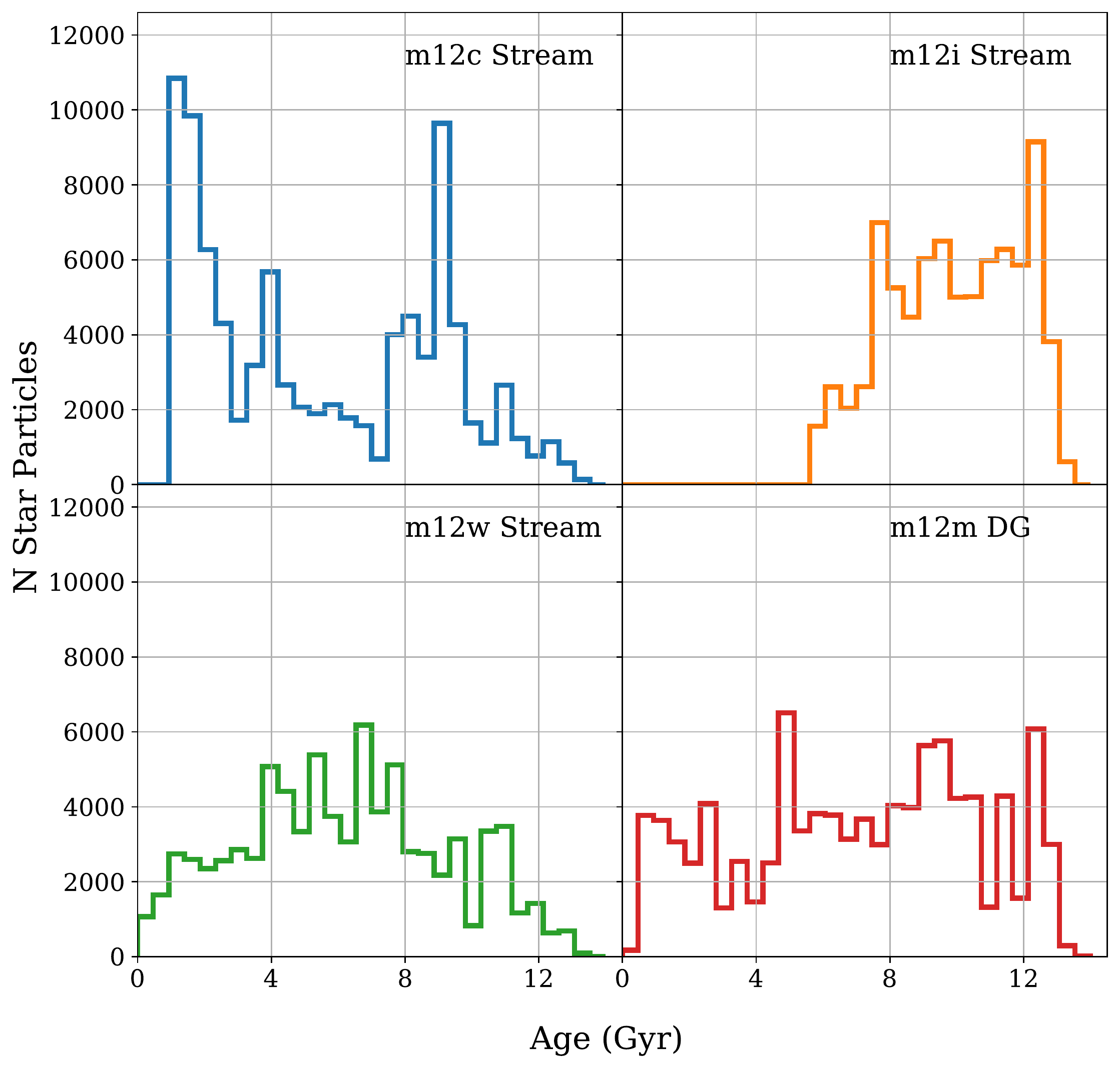}
    \caption{Age distributions of star particles for the massive streams and dwarf galaxy from the P21 catalog. The peaks in the age distributions for m12c illustrate that star formation history of the m12c stream is significantly burstier than the other massive streams and the massive dwarf galaxy, with a burst at $\sim 9$ Gyr and a burst around $\sim 2$ (following its first pericentric passage). The stream in m12c forms approximately $25\%$ of its mass after its first pericentric passage. As a result of its bursty star formation history, it does not enrich as steadily as the other massive systems.}
    \label{fig:sfh_massive}
\end{figure}

To get insight as to why m12c is more metal-poor than the other systems with comparable stellar mass, Figure \ref{fig:sfh_massive} shows the age distributions (i.e., the star formation histories) of the star particles in each system. The stream in m12c has a very peaky age distribution, indicating two major bursts of star formation. The first burst occurs at around $\sim 9$ Gyr, while the second burst occurs around $\sim 2$ Gyr. The stream progenitor first crosses the virial radius of the halo 3 Gyr ago, and experiences its first pericentric passage 1.5 Gyrs ago; its last burst of star formation, which is responsible for approximately $25\%$ of the stellar mass in the system, begins after the progenitor has entered the halo and continues beyond its first pericentric passage. This is in stark contrast to the framework in, for example, the \cite{Bullock2005} simulations, in which all dwarfs are assumed to lose their gas to ram pressure stripping upon falling into the main host halo. 

However, this is consistent with a recent study \citealt{DiCintio2021}, who find that close pericentric passages of relatively gas-rich systems can result in pericentric passage triggered starbursts, arising due to compression of cold gas. 
\cite{DiCintio2021} find that 25\% of the dwarfs in their simulations experience these starbursts, requiring that gas mass fraction be $M_{\rm gas,inf}/M_{\rm vir,inf} \geq 10^{-2}$ and pericenter be greater than 10 kpc. An example of gas compression at pericenter triggering star formation in satellites in the Auriga simulations is also discussed in \cite{Simpson2018}. While a detailed exploration of the gas properties of the halo progenitors in the \textit{Latte} simulations is beyond the scope of this work (see Samuel et al. 2021, in prep), we emphasize that, in these simulations, star formation after infall (and after first pericentric passage) can significantly contribute to the stellar mass of halo progenitors, and affect its overall CARD.

Figures \ref{fig:age_v_fe} and \ref{fig:sfh_massive} highlight both a limitation of the CARDS technique as well as the potential insights we can glean from studying the abundance patterns of halo progenitors. The diversity in the SFHs and the abundance distributions at high masses means that their properties are not necessarily well recovered from the templates; if the abundance distribution of the halo progenitor is not well represented by the template abundance distribution, the technique will not derive the correct parameters for the assembly history. However, signatures of these interesting formation histories are encoded in the abundance patterns. We therefore emphasize the fact that measuring stellar parameters and abundances within the MW stellar halo provides a unique window into galaxy formation in the high redshift universe.

\subsection{Future Work Within the Simulations}
\label{sec:future}
While there are limitations to using CARDs to constrain assembly histories, as discussed in the previous section, there are also prospects for further improving and expanding this method. First of all, a natural extension of the work presented here is to supplement the P21 catalog with the earlier accretion events identified and presented in Horta et al. (2021, in prep). The Horta catalog includes accreted galaxies down to a stellar mass limit of $10^7 M_{\odot}$; extending the catalog of known accretion events in the \textit{Latte} suite down to lower mass limits at earlier times is an ongoing effort. Once we have complete catalogs in hand for the progenitors of all the \textit{Latte} halos, it will be possible to test the effectiveness of the CARDs framework at recovering the full assembly histories of the \textit{Latte} suite.

In addition, improvements in the simulation may improve our ability to constrain the quenching time of our halo progenitors. The upcoming FIRE-3 physics model (Hopkins et al., in prep) includes many updates and improvements in modeling stellar evolution. In addition to the updated physics, there will also be a flexible new framework for exploring nucleosynthesis and metal enrichment (Wetzel et al., in prep), enabling the user to implement different feedback and chemical enrichment models in post-processing. With this model (known as the ``age-tracer" model), revised yields (as well as additional sources of enrichment, including r- and s- process) can be implemented to compute revised abundances at $z=0$. This has particularly exciting implications for this work: creating CARD templates that include information about more elements, including those that trace chemical enrichment on different timescales, are likely to improve the ability of the CARDs method to recover the quenching times of halo progenitors. Furthermore, the option for implementing flexible yields enables the option for tuning the $z=0$ abundance distributions to match observed dwarf galaxies. After matching the simulated CARDs to observed dwarf galaxy CARDs, we could create a template set from the simulations that could be used on the observations.

\section{Conclusions}
\label{sec:concl}

The halo of the MW contains the remnants from as many as thousands of low mass galaxies that formed at high redshift and did not survive to present-day. However, we currently lack quantitative frameworks for quantifying the contributions from these systems and identifying the stars that were born within them. In this paper, we have experimented with modeling the chemical abundance ratio distributions (CARDs) of the stellar halos of the \textit{Latte} simulations, as a linear combination of empirical templates for halo progenitors with different stellar masses and accretion times.  

In the Introduction, we pose three questions we seek to address in this work. We return to these questions here, and summarize our main results.

\begin{enumerate}
    \item \textit{Does the method proposed by L15 of modeling the stellar halo CARD as a linear combination of templates work in a more realistic setting?} \textbf{We find that CARDs can be effectively used to explore histories in a realistic setting.}
    \begin{itemize}
        \item We demonstrate the link between formation history and CARDs in the \textit{Latte} simulations. Each of the \textit{Latte} halos has distinct CARDs, as a result of their diverse formation histories.
        \item The CARDs method recovers the mass spectrum very accurately (to within 10\%) for four out of the seven presented here. The CARDs method shows promise for accurately inferring the mass spectrum of accreted dwarfs in LG galaxies.
        \item The fraction of mass accreted as a function of halo progenitor quenching time is generally inferred less accurately than the mass spectrum. We infer the accreted mass fraction as a function of progenitor quenching time with a precision of $20-30$\% for five out of the seven halos studied in this work. We discuss prospects for improving the performance of this technique, such as including information from additional elements, as well as upcoming improvements to the chemical evolution models in the simulations. 
        
    \end{itemize}
    \item \textit{What is an observationally viable method for constructing templates? Can we use the CARDs of present-day dwarf galaxies to infer the properties of the disrupted dwarfs?} \textbf{We present a method for using dwarf galaxies to construct templates for halo progenitors, and find that the templates from dwarf galaxies successfully recover the halo assembly histories.}
    \begin{itemize}
        \item Building off of the discussion of P21, we present the differences between the present day dwarf galaxies, coherent streams, and phase-mixed debris. At fixed stellar mass, dwarf galaxies tend to have longer $t_{100}$'s than coherent streams and phase-mixed debris. At fixed $t_{100}$, phase-mixed debris tends to have the largest stellar mass. This in turn has implications for the CARDs of the different populations: we show how phase-mixed debris is, on average, enhanced in magnesium relative to dwarf galaxies with the same mean metallicity. Therefore, a framework for using dwarf galaxy CARDs to model streams and phase-mixed debris must take this effect into account. 
        \item We present a method for creating templates for halo progenitors (i.e. streams or phase-mixed debris) using exclusively present-day dwarf galaxies, that accounts for the fact that present-day dwarf galaxy populations generally form their stars over a longer period of time. Our S/PM and DG templates are generally in good agreement.
        
    \end{itemize}
    \item \textit{How can we assess the accuracy of this method?} \textbf{We identify failure cases to this method when the best-fit models are poor fits to the data, as indicated by high model residuals.}
    \begin{itemize}
        \item For two out of the seven halos studied in this work, the CARDs from the templates (using the best fit model) are not good fits to the data, as demonstrated by the high RMS dispersion of the residuals. We find that a poor fit to the data generally indicates that the derived properties of the accretion histories are untrustworthy. This arises as a result of the fact that the abundance distributions for all dwarf galaxies and halo progenitors exhibit diversity in their star formation and chemical enrichment histories. We highlight an example due to pericentric passage induced starburst in m12c. Therefore, in interpreting the distributions of halo progenitors, it remains important to recall that we are studying the remains of galaxies, which may have experienced their own unusual histories before being disrupted in the MW halo. 
    \end{itemize}
\end{enumerate}

This work marks only a first step in exploring the full capabilities of using detailed studies of dwarf galaxy populations to unravel the MW assembly history using CARDs. One complication not addressed in this work includes observational uncertainties; the precision required for spectroscopic abundance measurements (for both halo stars and dwarf galaxy stars) remains to be explored. Furthermore, we have neglected the presence of an in-situ halo population. 
In addition, we have constructed our templates in the context of simulations, where we have had access to all information about our dwarf galaxies, including ages and star formation histories; the optimal methods for constructing dwarf galaxy templates from observations remains to be explored. 

However, in this work, we have undertaken a critical first step of testing the CARDs modeling framework in the context of a cosmological simulation. We emphasize again that the model presented here contains no physics or prior information, and there are many ways in which more complexity and information could be incorporated into this model. For example, prior information about the mass of the halo, the mass function of DM halos from simulations, or kinematic information are all likely to improve the quality of the inferences. However, for the purposes of this paper, we demonstrate the wealth of information that can be extracted from chemical abundances alone. The power of chemical abundances alone is especially important for characterizing the origin of the phase-mixed halo.

Through the current and next generation of spectroscopic surveys, we are well positioned to use halo star chemical abundances to learn more about the evolutionary histories of the halo progenitors of the MW and M31. Only in the LG can we test our understanding of star formation and chemical enrichment in low mass galaxies in the high redshift universe, on a star by star basis.

\acknowledgements{ECC is supported by a Flatiron Research Fellowship at the Flatiron Institute. The Flatiron Institute is supported by the Simons Foundation. The data used in this work were, in part, hosted on facilities supported by the Scientific Computing Core at the Flatiron Institute, a division of the Simons Foundation. ECC would like to thank Duane Lee, Alis Deason, Erin Kado-Fong, Ted Mackereth, and the members of the CCA Dynamics Group and Astronomical Data Group for useful scientific discussions.
RES and NP acknowledge support from NASA grant 19-ATP19-0068. RES further acknowledges support from NSF grant AST-2009828 and grant HST-AR-15809 from the Space Telescope Science Institute (STScI), which is operated by AURA, Inc., under NASA contract NAS5-26555. KVJ's contributions were supported by NSF grant AST-1715582.
AW received support from: NSF grants CAREER 2045928 and 2107772; NASA ATP grants 80NSSC18K1097 and 80NSSC20K0513; HST grants GO-14734, AR-15057, AR-15809, GO-15902 from STScI; a Scialog Award from the Heising-Simons Foundation; and a Hellman Fellowship. IE is supported by a Carnegie-Princeton Fellowship through the Carnegie Observatories. CAFG was supported by NSF through grants AST-1715216, AST-2108230,  and CAREER award AST-1652522; by NASA through grant 17-ATP17-0067; by STScI through grant HST-AR-16124.001-A; and by the Research Corporation for Science Advancement through a Cottrell Scholar Award.
We ran simulations using: XSEDE, supported by NSF grant ACI-1548562; Blue Waters, supported by the NSF; Pleiades, via the NASA HEC program through the NAS Division at Ames Research Center.
}

\bibliography{refs}

\begin{appendix}
In this appendix, we present additional figures and analysis for the interested reader. In Section \ref{app:temps}, we present example template sets in [Mg/C] vs [Fe/H] used in this analysis. In Section \label{app:detailed_res}, we present detailed results for all the remaining \textit{Latte} halos studied in this work. As in Section \ref{sec:results}, we present the model CARDs for each halo as well as the inferred mass spectrum and fraction of mass accreted as a function of progenitor quenching time. In Section \ref{app:lower_mass}, we present results for modeling the \textit{Latte} halos excluding halo progenitors with stellar masses greater than $10^8 M_{\odot}$.

\section{Additional Templates}
\label{app:temps}

\begin{figure}
    \centering
    \includegraphics[width=0.8\textwidth]{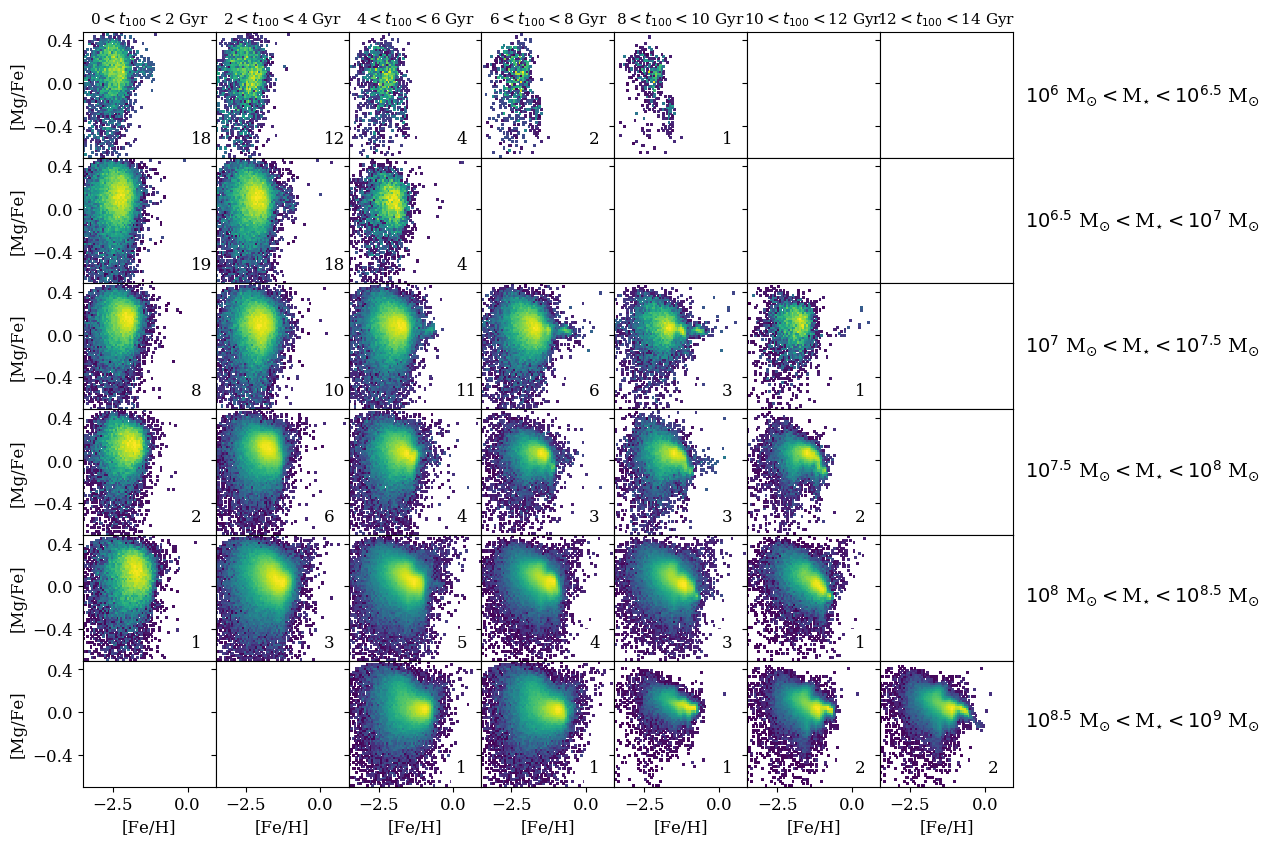}
    \includegraphics[width=0.8\textwidth]{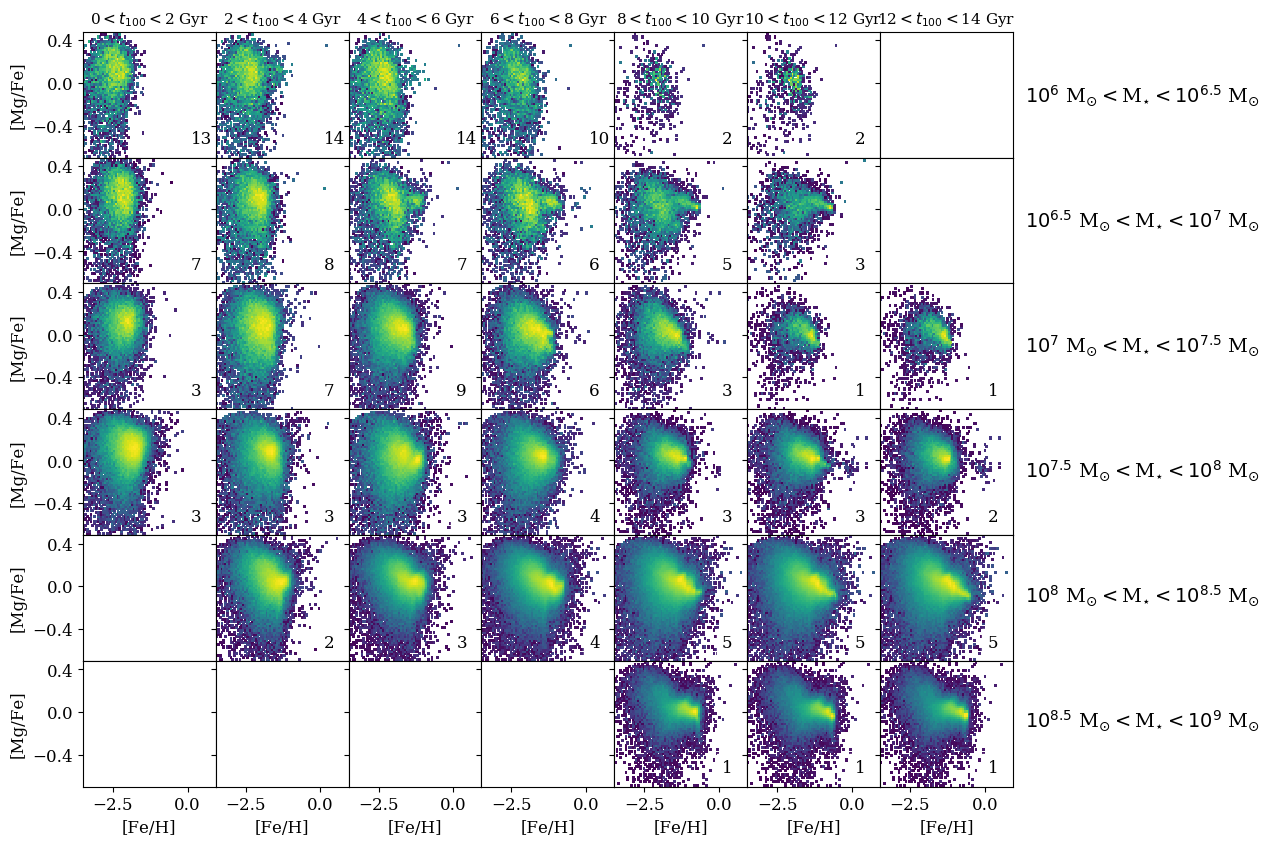}
    \caption{CARD templates in [Mg/C] vs [Fe/H]. Top panels show the templates constructed from the streams and phase-mixed debris, while the lower panels show the templates constructed from present-day dwarf galaxies.}
    \label{fig:mgc_temps}
\end{figure}

Figure \ref{fig:mgc_temps} shows the [Mg/C] vs [Fe/H] templates (referred to as the [Mg/C] Templates throughout the text), constructed from streams and phase-mixed debris (top) and present-day dwarf galaxies (bottom). In this work, we found that including information from carbon marginally improved the inferred mass spectrum (see Section \ref{sec:concl}); however, we have dealt exclusively with templates constructed in two dimensions. We could further experiment with constructing templates with additional dimensions. 

While we don't show them here, we also mention that we experimented with constructing ``master" template sets, constructed from streams, phase-mixed debris, and dwarf galaxies. However, we didn't include the results from this analysis here, as the results were essentially identical to the results from the DG templates. With the exception of the most massive row of templates, combining the S/PM sources with the DG sources did not significantly change the average template densities, resulting in very similar model CARDs to those produced by the DG templates. 

For a subset of the halos, we also experiment modeling their early accretion events (accreted prior to 6 Gyr ago) using only an ``early" template set (constructed from sources with $t_{100}<6 Gyr$ and a ``late" template set (from sources with $t_{100}>6 Gyr$). For the halos we tested, we find that they perform comparably, as with the S/PM and DG templates. For one of the halos we tested (m12f), the ``late" templates did perform much better than the early templates. Because m12f experiences the most massive accretion event prior to 6 Gyr, its CARD is better matched by the older stellar populations of massive streams and dwarf galaxies with later $t_{100}$'s. Therefore, as with the other template sets, if the stellar populations of the halo are well represented by the templates, we will do a reasonable job of inferring the assembly history. 

\section{Detailed Results for All Simulations}
\label{app:detailed_res}

\begin{figure}
    \centering
    \includegraphics[width=\textwidth]{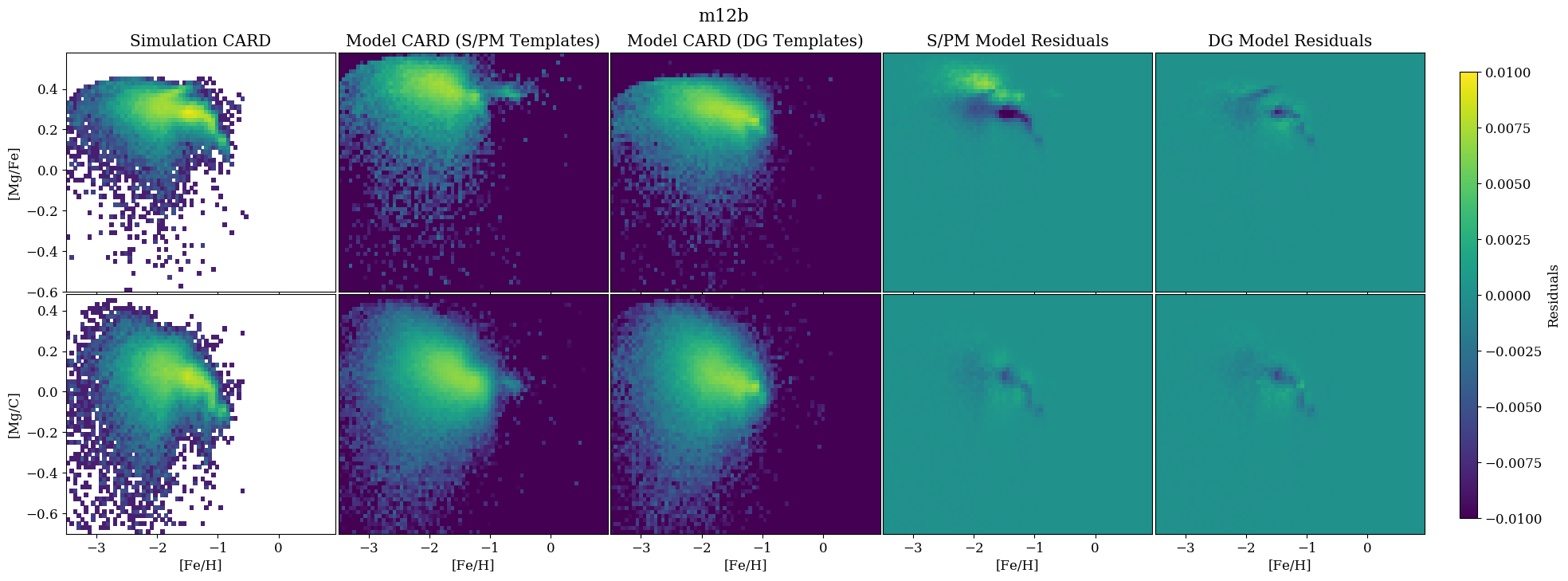}
    \includegraphics[width=\textwidth]{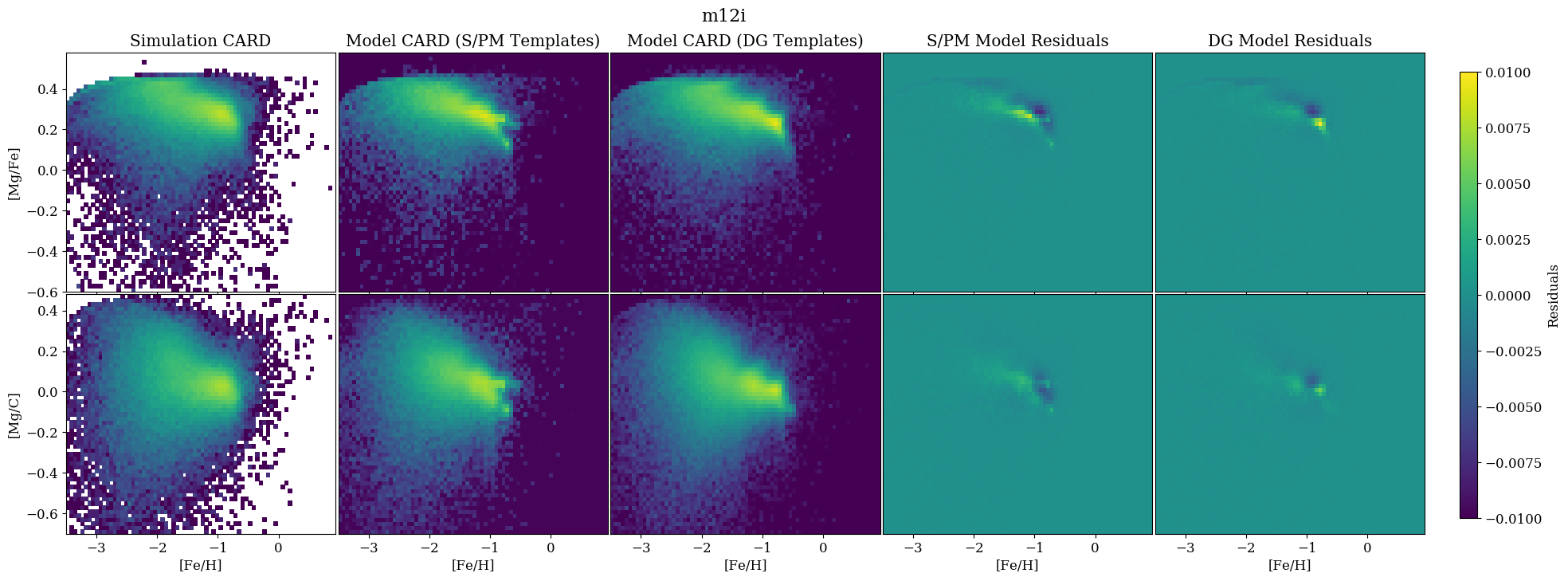}
    \includegraphics[width=\textwidth]{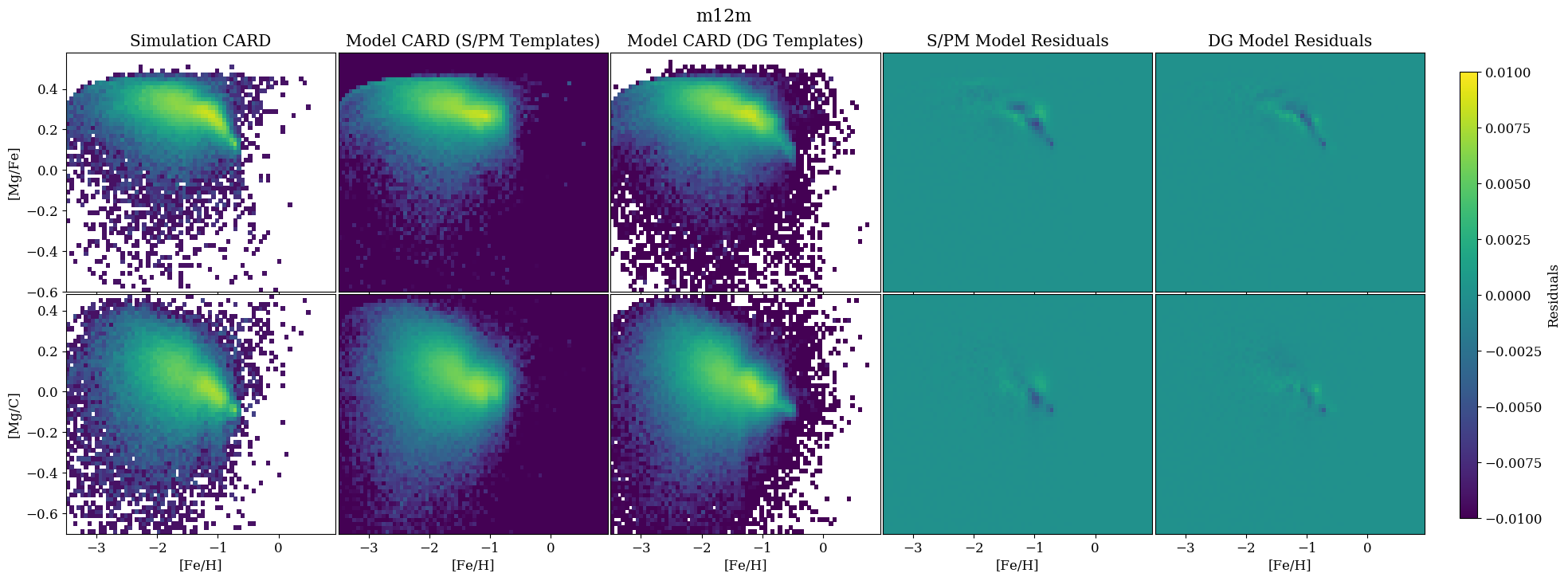}
    \caption{Same as Figures \ref{fig:m12f_dens} and \ref{fig:m12c_dens}, but for the simulations m12b, m12i, and m12m.}
    \label{fig:m12bim_dens}
\end{figure}

\begin{figure}
    \centering
    \includegraphics[width=\textwidth]{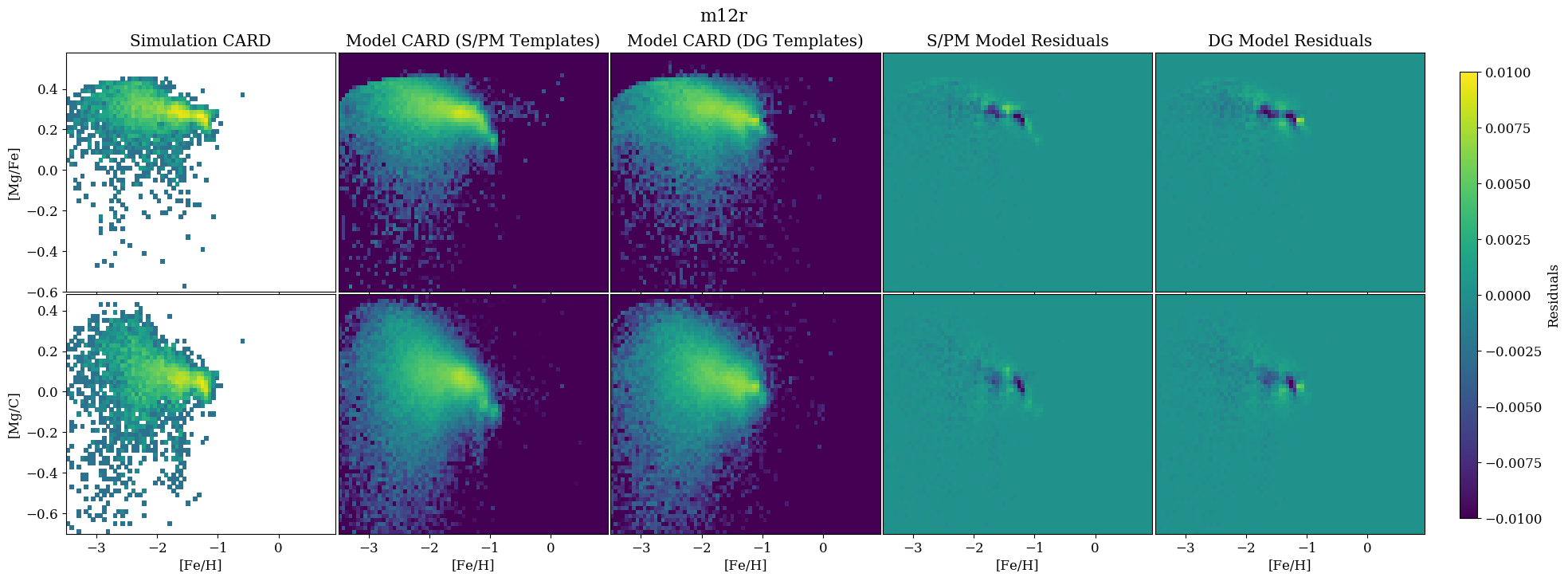}
    \includegraphics[width=\textwidth]{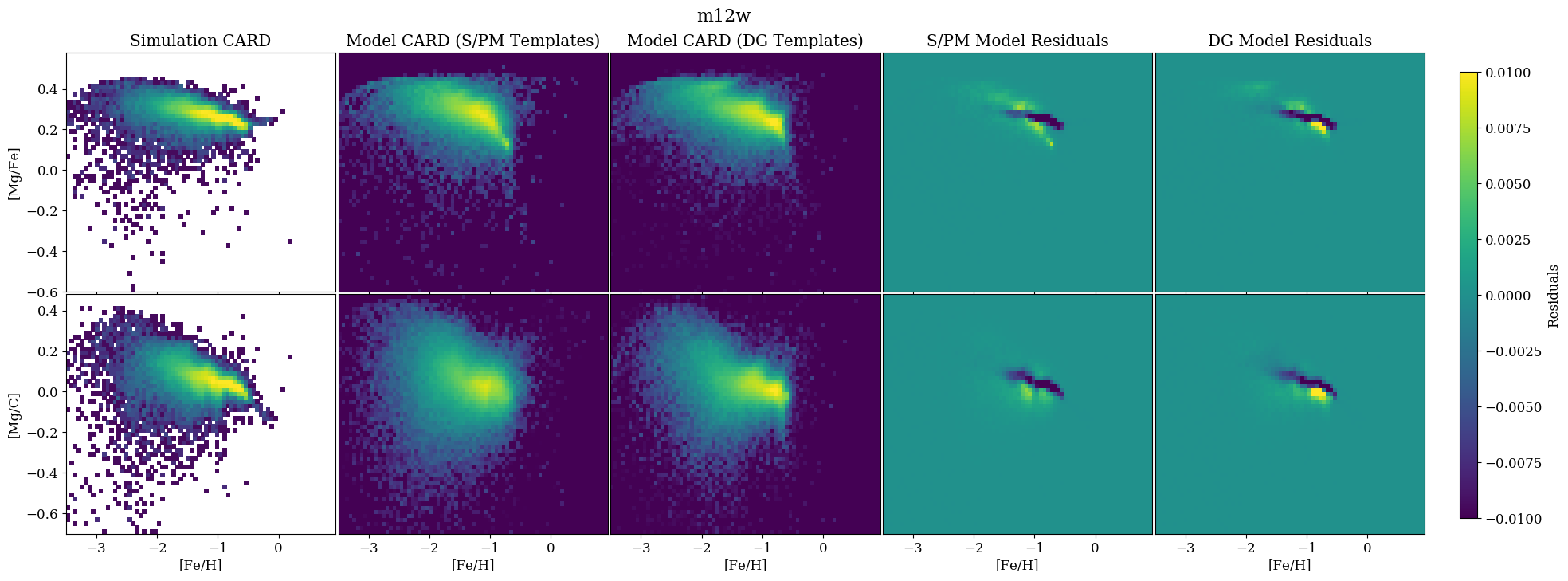}
    \caption{Same as Figures \ref{fig:m12f_dens}, \ref{fig:m12c_dens}, and \ref{fig:m12bim_dens} but for the simulations m12r and m12w.}
    \label{fig:m12rw_dens}
\end{figure}

\begin{figure}
    \centering
    \includegraphics[width=\textwidth]{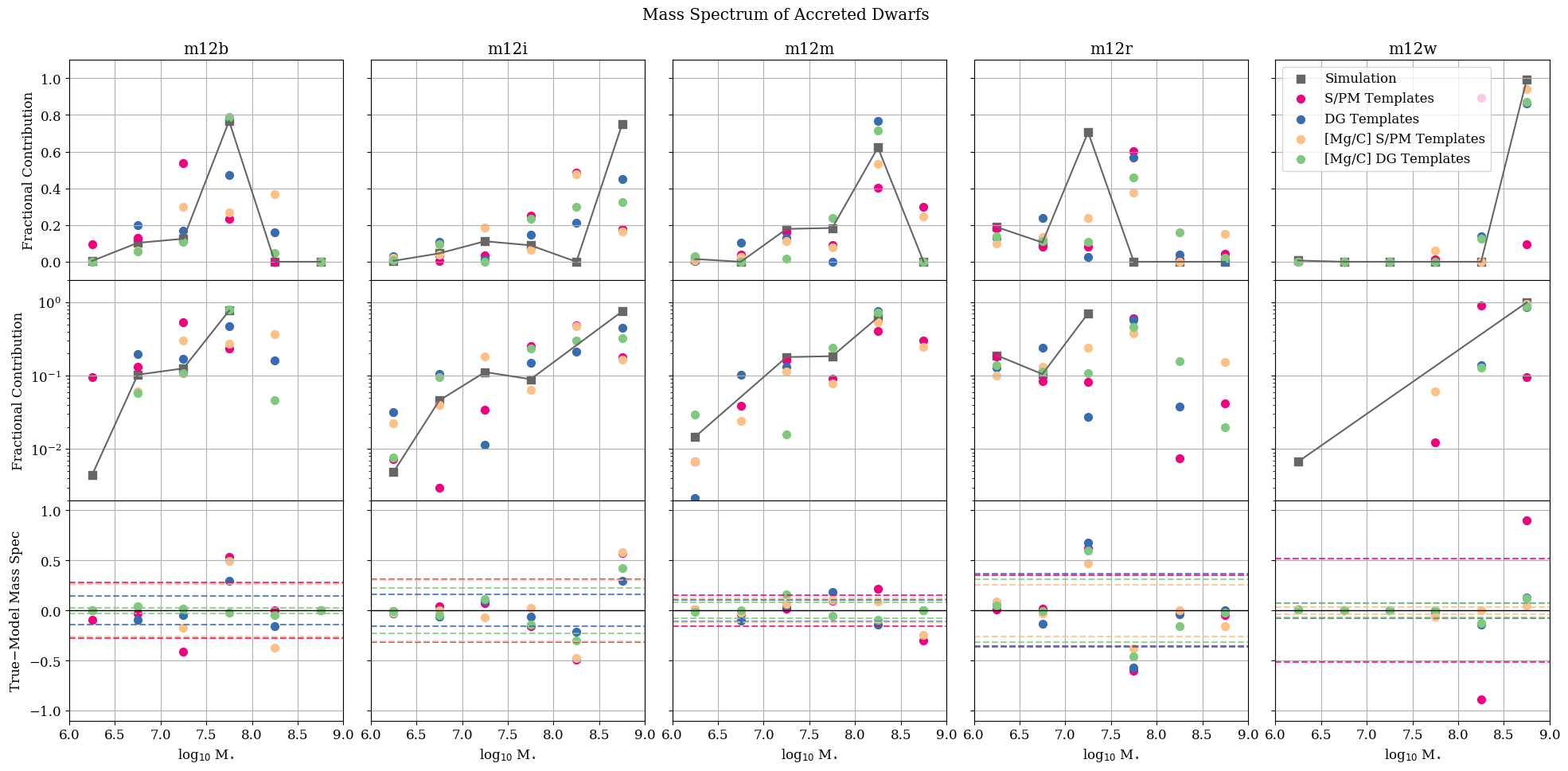}
    \caption{Same as Figure \ref{fig:m12fc_mass}, but for the remaining \textit{Latte} stellar halos not discussed in detail in Section \ref{sec:results}. As in previous figures, grey squares mark the true values from the simulations, while pink, blue, peach and green circles show the results from the S/PM templates, the DG templates, the [Mg/C] S/PM templates, and the [Mg/C] DG templates, respectively. }
    \label{fig:mass_spec_app}
\end{figure}

\begin{figure}
    \centering
    \includegraphics[width=\textwidth]{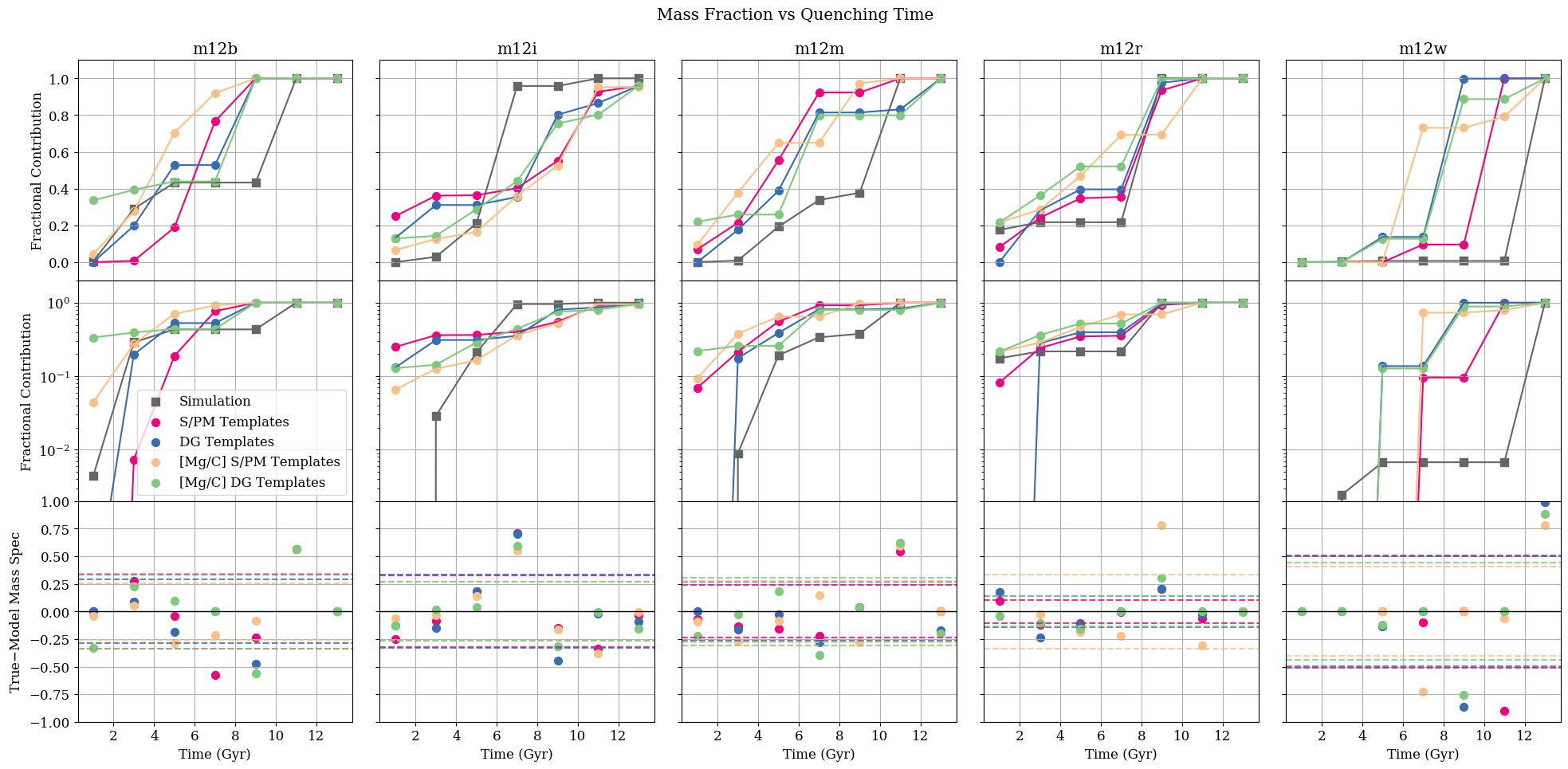}
    \caption{Same as Figure \ref{fig:m12fc_time}, but for the remaining \textit{Latte} stellar halos not discussed in detail in Section \ref{sec:results}. As in previous figures, grey squares mark the true values from the simulations, while pink, blue, peach and green circles show the results from the S/PM templates, the DG templates, the [Mg/C] S/PM templates, and the [Mg/C] DG templates, respectively. }
    \label{fig:mass_time_app}
\end{figure}

In this appendix, we present and describe the detailed results from the five other simulations studied in this work. In Figures \ref{fig:m12bim_dens}--\ref{fig:mass_time_app} we show results from all the remaining simulations; we discuss the failure cases of this method in more detail below. 

\begin{itemize}
    \item \textbf{m12i:} The second column of panels in Figures \ref{fig:mass_spec_app} and \ref{fig:mass_time_app} highlight that none of the models do particularly well at recovering the assembly history of m12i. The reason for this is again the fact that the most massive accretion event in m12i ($M_{\star}=, t_{100}=$) is not represented by the templates. We can see this if we look back to Figure \ref{fig:stpm_tmps}: the most massive accretion event from m12i is the only accretion event that contributes to the third and fourth panels along the bottom row of the S/PM templates in Figure \ref{fig:stpm_tmps}. Furthermore, we can these squares are left blank in Figure \ref{fig:dg_tmps}, as there are no dwarf galaxies included in P21 whose stellar populations can be used to populate those grid squares. This event is the only one within the entire P21 catalog to have $M_{\star}>10^{8.5}$ and a quenching time $t_{100}<8$ Gyr. Therefore, the templates constructed here do not accurately recover its properties. Expanding the templates to include more massive systems could help improve the performance of the method here.
    \item \textbf{m12r:} In Figure \ref{fig:form_age}, we highlighted the formation history of m12r characterized by massive, recent accretion. From Figure \ref{fig:form_age_acc_parts}, it becomes clear these massive, recent accretion events are too massive to make it into the P21 catalog; the top left panels of Figure \ref{fig:m12rw_dens} show the CARD for the m12r particles modeled in this work. This abundance distribution looks different from the other halos, and has a feature at high metallicity and relatively high magnesium that is inconsistent with the other halos and the templates. This feature results in poorly inferred mass spectrum (Figure \ref{fig:mass_spec_app}), making m12r one of the failure cases for this method. 
    
    The reason that m12r has a strange abundance distribution is a result of group infall. While the most massive infalling satellites are excluded from this analysis as a result of the P21 selection criteria, the dominant accretion event included in this analysis in m12r ($M_{\star}=1.9 \times 10^{7} M_{\odot}, t_{100}=9.1$ Gyr) has a CARD that extends to higher metallicity than its counterparts in the template sets. This halo progenitor falls in with one of the massive infalling satellites ($M, t_{100}$), experiencing several pericentric passages during their collective infall and bursts of star formation during this interaction. Because it has a higher metallicity than expected based on its stellar mass and $t_{100}$, the CARDs method infers a stellar mass that is too high for this progenitor. While this is technically a ``failure case" for the method, it is also another example of how clues as to the interesting evolutionary histories of the halo progenitors are encoded in the halo CARDs. 
    \item \textbf{m12w:} Results from modeling the stellar halo of m12w are shown in the lower panels of \ref{fig:m12rw_dens}. The simulation m12w is an interesting case, as its mass spectrum is inferred extremely accurately, but the timing of its major accretion event is inferred inaccurately, and the CARD models show relatively high residuals for all models. From the residuals, we see that the shape of the turnover in the [Mg/Fe] vs [Fe/H] plane is not a good match to the templates; however, as can be seen in Figure \ref{fig:age_v_fe}, the iron distribution as a function of age for the m12w massive stream isn't too dissimilar from that of the massive dwarf in m12m. Therefore, while there are high residuals as a result of the different Mg distributions, the iron distributions are similar enough such that the mass of the accretion event is recovered accurately wit the DG templates (though not its quenching time). 
\end{itemize}

\section{Modeling Only Lower Mass Events}
\label{app:lower_mass}

In Section \ref{sec:discussion}, we discussed how diverse evolutionary histories, particularly for the massive halo progenitors, resulted in failure modes for the CARDs method. In this appendix, we repeat our analysis, but excluding all accretion events above $M_{\star}>10^{8} M_{\odot}$. Given we can often use kinematic information in conjunction with chemical information to identify stars belonging to the most massive accretion events (e.g., Sagittarius, Gaia-Enceladus-Sausage), we test here the performance of the CARDs method once the most massive progenitors are subtracted off. 

\begin{figure}
    \centering
    \includegraphics[width=\textwidth]{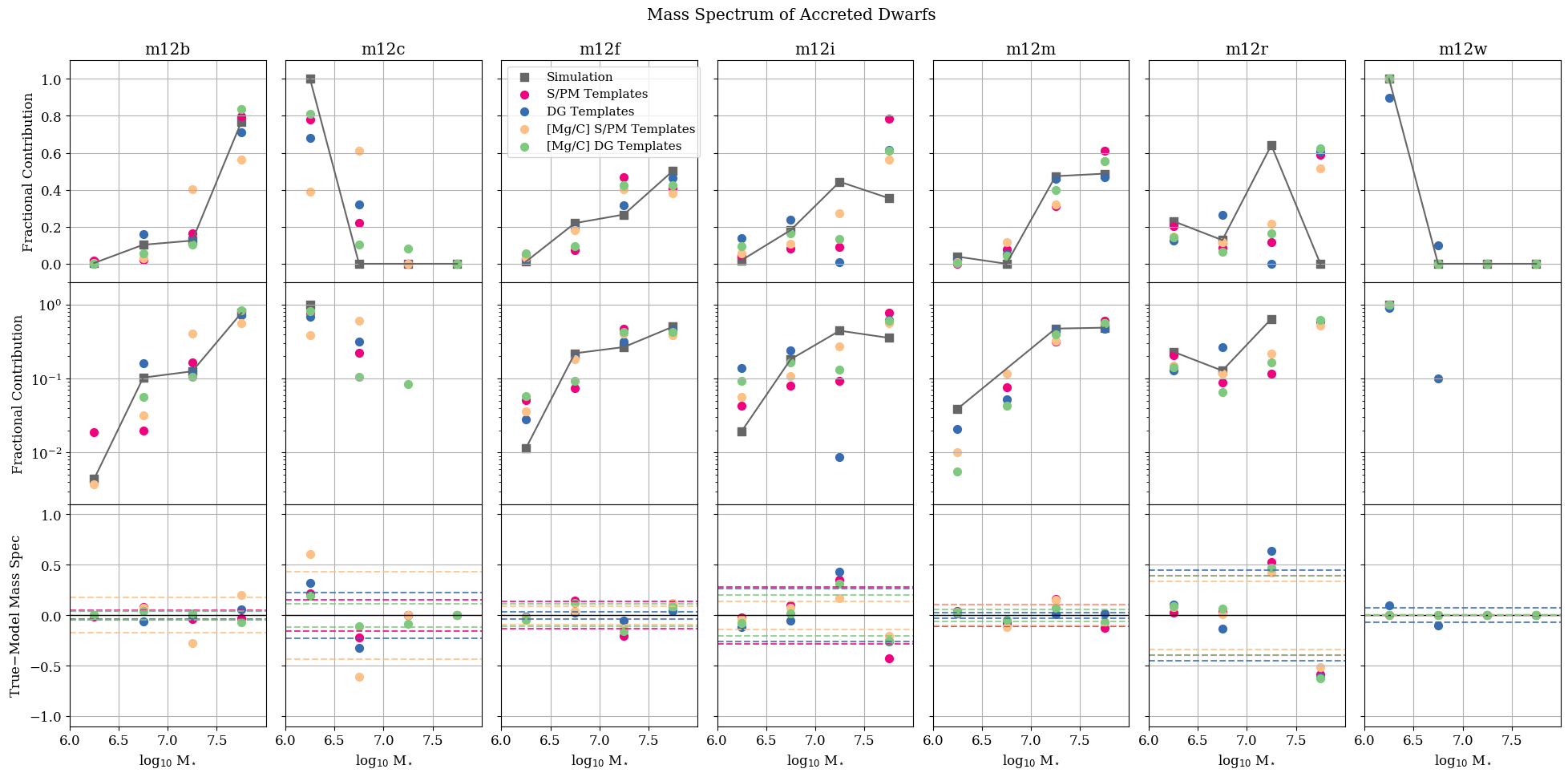}
    \caption{Same as Figure \ref{fig:mass_spec_app}, but excluding accretion events above $M_{\star}>10^8 M_{\odot}$. As in previous figures, grey squares mark the true values from the simulations, while pink, blue, peach and green circles show the results from the S/PM templates, the DG templates, the [Mg/C] S/PM templates, and the [Mg/C] DG templates, respectively.}
    \label{fig:lower_mass_spec}
\end{figure}

Figure \ref{fig:lower_mass_spec} shows the resulting inferred mass spectra, over the stellar mass range of $10^{6} M_{\odot}<M_{\star}<10^{8} M_{\odot}$. With the exception of m12r, whose results are unchanged (there is no accretion event with $M_{\star}>10^{8} M_{\odot}$ included in m12r in the P21 catalog), we see improvements in the residuals for the inferred mass spectra for all halos. 

\begin{figure}
    \centering
    \includegraphics[width=\textwidth]{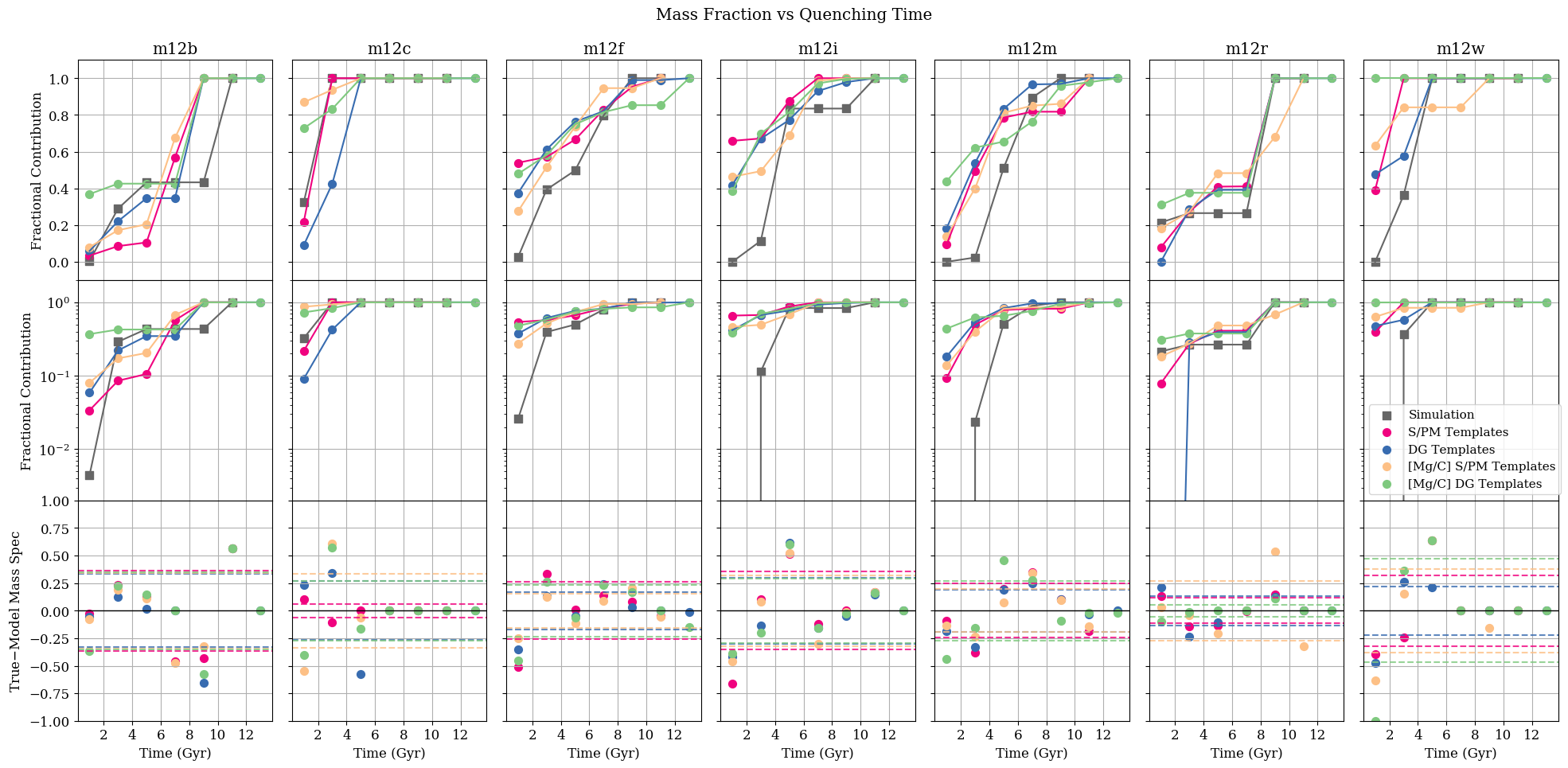}
    \caption{Same as Figure \ref{fig:mass_time_app}, but excluding accretion events above $M_{\star}>10^8 M_{\odot}$. As in previous figures, grey squares mark the true values from the simulations, while pink, blue, peach and green circles show the results from the S/PM templates, the DG templates, the [Mg/C] S/PM templates, and the [Mg/C] DG templates, respectively. }
    \label{fig:lower_mass_time}
\end{figure}

\end{appendix}
\end{document}